\documentclass[journal,11pt,onecolumn]{IEEEtran}

\newcommand{\be}{\begin{equation}}
	\newcommand{\ee}{\end{equation}}
\newcommand{\ben}{\begin{equation*}}
	\newcommand{\een}{\end{equation*}}
\newcommand{\ba}{\begin{array}}
	\newcommand{\ea}{\end{array}}
\newcommand{\Z}{{\mathcal Z}}
\newcommand{\A}{{\mathcal A}}
\newcommand{\Y}{{\mathcal Y}}

\newcommand{\F}{{\mathcal F}}

\newcommand{\defi} { \stackrel{\bigtriangleup}{=} }

\newtheorem{theorem}{Theorem}

\newtheorem{remark}{Remark}

\newtheorem{proposition}{Proposition}

\usepackage{bibspacing}
\usepackage{balance}
\usepackage{cite}

\usepackage{graphicx}
\usepackage[cmex10]{amsmath}
\usepackage{enumerate}

\usepackage{subfig}
\usepackage[latin1]{inputenc}

\usepackage{color}
\usepackage{colortbl}

\usepackage{multirow}

\usepackage{verbatim}

\usepackage{epstopdf}
\usepackage{multirow}
\usepackage{amssymb}

\usepackage[mathscr]{euscript}
\usepackage{hyperref}
\usepackage{float}

\begin{document}

\title{Joint Attack Detection and Secure State Estimation of Cyber-Physical Systems}

\author{\IEEEauthorblockN{Nicola Forti\IEEEauthorrefmark{2}\IEEEauthorrefmark{1},
		Giorgio Battistelli\IEEEauthorrefmark{1}, Luigi Chisci\IEEEauthorrefmark{1}, and
		Bruno Sinopoli\IEEEauthorrefmark{3}} \\
	\vspace{.2cm}	
	
	\IEEEauthorrefmark{2}Research Department, NATO STO CMRE, La Spezia, Italy \\
	\IEEEauthorrefmark{1}Dipartimento di Ingegneria dell'Informazione, Universit\`{a} di Firenze, Firenze, Italy \\
%    \IEEEauthorrefmark{3}Department of Electrical and Computer Engineering, Carnegie Mellon University, Pittsburgh, PA \\
\IEEEauthorrefmark{3}Department of Electrical \& Systems Engineering, Washington University in St. Louis, MO \\
	\vspace{.2cm}
	
        \IEEEauthorrefmark{2}nicola.forti@cmre.nato.int,
		\IEEEauthorrefmark{1}\{giorgio.battistelli,luigi.chisci\}@unifi.it,
%		\IEEEauthorrefmark{3}brunos@ece.cmu.edu
\IEEEauthorrefmark{3}bsinopoli@wustl.edu
}

%\corres{*Nicola Forti, Research Department, 
%	NATO-STO CMRE, \\
%%	NATO-STO Centre for Maritime Research and Experimentation, \\
%	viale San Bartolomeo 400, 19126 La Spezia, Italy. \\
%	\email{nicola.forti@cmre.nato.int}}
\maketitle

\abstract 
This paper deals with secure state estimation of cyber-physical systems subject to switching (on/off) attack signals and injection of fake packets (via either packet substitution or insertion of extra packets).
The random set paradigm is adopted in order to model, via \emph{Random Finite Sets} (RFSs), the switching nature of both system attacks and the injection of fake measurements.
The problem of detecting an attack on the system and jointly estimating its state, possibly in the presence of fake measurements, is then formulated and solved in the Bayesian framework
for systems with and without direct feedthrough of the attack input to the output.
This leads to the analytical derivation of a \emph{hybrid Bernoulli filter} (HBF) that updates in real-time the joint posterior density of a Bernoulli attack RFS and of the state vector.
A closed-form Gaussian-mixture implementation of the proposed hybrid Bernoulli filter is fully derived in the case of invertible direct feedthrough.
Finally, the effectiveness of the developed tools for joint attack detection and secure state
estimation is tested on two case-studies concerning
a benchmark system for unknown input estimation and a standard IEEE power network application.


\section{Introduction}\label{sec1}

Cyber-Physical Systems (CPSs) are complex engineered systems arising from the integration of computational resources and physical processes, tightly connected through a communication infrastructure. 
Typical examples of CPSs include next-generation systems in building and environmental monitoring/control, health care, electric power grids, transportation and mobility and industrial process control. 
While, on one hand, advances in CPS technology will enable enhanced autonomy, efficiency, seamless interoperability and cooperation, on the other hand the increased interaction between cyber and physical realms is unavoidably providing novel 
security vulnerabilities, which make CPSs subject to non-standard malicious threats.
Recent real-world attacks such as the Maroochy Shire sewage spill, the Stuxnet worm sabotaging an industrial control system, and the 
%German steel mill damage 
lately reported massive power outage against Ukrainian electric grid \cite{ICS},
 have brought into particularly sharp focus the urgency of designing secure CPSs. 
%%%%%%%%%%%%%%%%%%%%%%%% STATE OF ART
It is worth pointing out that in presence of malicious threats against CPSs, standard approaches extensively used for control systems subject to benign faults and failures are no longer suitable.
Moreover, the design and implementation of defense
mechanisms usually employed for cyber security, can only guarantee limited layers of protection, since they do not take into account vulnerabilities like the ones on physical components.
This is why recent research efforts on the design of secure systems have explored different routes. 
Preliminary work addressed the issues of attack detection/identification, and 
proposed attack monitors for deterministic control systems \cite{Pasqualetti2013}. 
Secure strategies have been studied for \textit{replay} attacks \cite{Mo2009,Miao13} where the adversary first records and then replays the observed data, as well as for \textit{denial-of-service} (DoS) attacks \cite{Tesi15,Zhang15}
disrupting the flow of data.
Moreover, active detection methods have been designed in order to detect \textit{stealthy} attacks via manipulation of, e.g., control inputs \cite{MoWee2015} or dynamics \cite{Sean2015}. 
%or robust detection of a binary random variable \cite{Mo2014}. 
Over the last few years, the problem of secure state estimation, i.e. capable of reconstructing the state even when the CPS of interest is under attack,
has gained considerable attention \cite{MoSi2015,Fawzi2014,Pajic17,Shoukry2015,Mishra17,Chong15,Teixeira2015,Shi17,Forti2016,TAC}.
%\cite{MoSi2015,TAC}. 
Initial work considered a worst-case approach for the special class of SISO systems \cite{MoSi2015}. 
Under the assumption of linear systems subject to an unknown but bounded number of \textit{false-data injection} attacks on sensor outputs, the problem for a noise-free system has been cast into an $\ell_0-$optimization problem, which can be relaxed as a more efficient convex problem \cite{Fawzi2014}, and, in turn, adapted to systems with bounded noise \cite{Pajic17}. 
Further advances tried to tackle the combinatorial complexity of the problem by resorting to satisfiability modulo theories \cite{Shoukry2015} 
and investigated, in the same context, the case of Gaussian measurement noise \cite{Mishra17} and the concept of observability under attacks \cite{Chong15}. 
Most recently, deterministic models of the most popular attack policies have been presented based on adversary's resources and system knowledge \cite{Teixeira2015},  
%and resilient strategies have been also proposed for noisy systems with direct feedthrough under both data injection and switching mode attacks \cite{Yong2015}. 
%Most recently, in 
and secure state estimation of CPSs has been addressed \cite{Shi17} by modeling in a stochastic framework the attacker's 
decision-making 
by assuming Markov (possibly uninformative) decision processes instead of unknown or worst-case models.

Though the literature on attack-resilient state estimation is quite abundant, most of the existing contributions have adopted a deterministic (worst-case) approach and/or have been restricted  to linear systems.
In practice, the system monitor (defender) might have some (even no) probabilistic prior knowledge on the attacker's strategy and the CPS of interest might easily be affected by nonlinearities.
In this respect, a Bayesian approach where prior knowledge on the attacks is characterized in terms of probability distributions and nonlinearities are possibly handled by particle filtering
or Gaussian-mixture methods, seems well suited and will be pursued in this paper. 
This allows great flexibility in that knowledge available to the attack monitor can range from complete knowledge to no prior knowledge (uninformative prior) depending on the assumed distributions.

%%%%%%%%%%%%%%%%%%%%%%%% INTRO ATTACK MODEL

Specifically, in this paper three different types of adversarial attacks on CPSs are considered:
(i) \textit{signal} attack, i.e. signal of arbitrary magnitude and location injected (with known structure) to corrupt sensor/actuator data, 
(ii) \textit{packet substitution} attack,  
describing an intruder that possibly intercepts and then replaces the system-generated measurement with a fake (unstructured) one, 
and (iii) \textit{extra packet injection}, a new type of attack against state estimation, already introduced in information security \cite{Gu2005,Zhang2007}, in which multiple counterfeit observations (junk packets) are possibly added to the system-generated measurement.
Note that the key feature distinguishing signal attacks on sensors from packet substitution, relies on the fact that the former are assumed to alter the measurement through a given structure (i.e., known measurement function), whereas the latter mechanism captures integrity attacks that spoof sensor data packets with no care of the model structure.
By considering both structured and unstructured injections, we do not restrict the type of attack the adversary can enforce on the sensor measurements.
Please notice that,  as a further by-product, the Bayesian approach with uninformative prior can also deal with the situation  in which the attacker has the ability to choose arbitrarily large attack and/or fake measurements, while the worst-case attack paradigm in this case is not viable.

The present paper aims to address the problem of simultaneously detecting a signal attack while estimating the state of the monitored system, possibly in presence of fake measurements independently injected into the system's monitor by cyber-attackers.
A random set attack modeling approach is undertaken by representing the signal attack presence/absence by means of a
Bernoulli random set (i.e. a set that, with some probability, can be either empty or a singleton depending on the presence or not of the attack) and by taking into account possible fake measurements by means of a random 
%%%measurement set. 
%%%According to (ii) and (iii), 
%%%the measurement set is represented by a Bernoulli or Poisson random set for the \textit{packet substitution} or, respectively, \textit{extra packet injection} attack.
measurement set. % represented by a Bernoulli or Poisson random set for the \textit{packet substitution} or, respectively, \textit{extra packet injection} attack.
We follow the approach of Forti et al. \cite{Forti2016},\cite{TAC} and formulate the  joint attack detection-state estimation problem  within the Bayesian framework as the recursive
determination of the joint posterior density of the signal attack Bernoulli set and of the state vector at each time given all the measurement sets available up to that time.
Strictly speaking, the posed Bayesian estimation problem is neither standard \cite{Ho1964} nor \textit{Bernoulli} filtering 
\cite{Ristic2013,Mahler2007,Vo2011,Vo2012} but is rather a \textit{hybrid} Bayesian filtering problem that aims to jointly estimate a Bernoulli random set
for the signal attack and a random vector for the system state.   
An analytical solution of the hybrid filtering problem has been found in terms of integral equations that generalize the Bayes and Chapman-Kolmogorov equations of the Bernoulli filter.
In particular, the proposed \textit{hybrid Bernoulli Bayesian filter} for joint attack detection-state estimation propagates in time,
via a two-step prediction-correction procedure, a joint posterior density completely characterized by a triplet consisting of:
(1) a signal attack probability; (2) a \textit{probability density function} (PDF) in the state space for the system under no signal attack; (3) a PDF in the joint attack input-state space for the system under signal attack.

The adopted approach enjoys the following positive features: 
1) it encompasses in a unique framework different types of attacks (signal attacks, packet substitution, extra packet injection, temporary DoS, etc.); 
2) it takes into account the presence of disturbances and noise and deals with general nonlinear systems;  
3) it propagates probability distributions of the system state, attack signal and attack existence, which can be useful for, respectively, real-time dynamic state estimation, attack reconstruction and security decision-making.
Notice that, unlike most previous work cited above, in the present paper we address the problem from the estimator's perspective and, hence, we cannot assume any specific strategy
for the attacker. This motivates the modeling of the signal attack as a switching unknown input affecting the system. 

%As compared to previous work by Forti et al., \cite{TAC} the main contributions of the present paper are: 
%(i) the extension of the model so as to include also packet substitution attacks;
%(ii) the derivation of a closed-form Gaussian-mixture hybrid Bernoulli filter for linear-Gaussian models.
Preliminary work on Bayesian state estimation against switching unknown inputs and extra packet injection was carried out by Forti et al. \cite{Forti2016},\cite{TAC}.
The present paper extends this preliminary work in the following directions.
\begin{enumerate}
\item It also considers the \textit{packet substitution} attack (in addition to the already considered \textit{extra packet injection} attack).
This novel type of attack refers to the practically relevant situation wherein the attacker has the ability to intercept and manipulate packets sent to the system monitor so as to replace system-originated measurements by fake ones but, unlike the extra packet injection attack, cannot send additional indistinguishable packets containing fake measurements to confuse the system monitor.
\item It provides the full derivation of a closed-form solution of the posed Bayesian filtering problem for linear-Gaussian models based on a Gaussian-mixture approach.
This allows a computationally efficient implementation of the proposed joint \textit{attack detector-state estimator} also extendable to nonlinear models via extended or unscented (instead of standard) Kalman filtering techniques.
\item It considers also the case of no direct feedthrough of the attack input into the observed output.
\end{enumerate}

The rest of the paper is organized as follows.
Section~\ref{sec2} introduces the considered attack models and provides the necessary background on joint input-and-state estimation as well as on random set estimation.
Sections~\ref{sec3} and \ref{sec4} formulate and solve the joint \textit{attack detection-state estimation} problem of interest in the Bayesian framework. 
%(proofs are omitted due to lack of space).
Section~\ref{sec5} provides detailed derivations of the Gaussian-mixture hybrid Bernoulli filer for linear-Gaussian models.
Then, Section~\ref{sec6} demonstrates the effectiveness of the proposed approach via numerical examples.
Finally, Section~\ref{sec7} ends the paper with concluding remarks and perspectives for future work.

\section{Problem Setup and Preliminaries}\label{sec2}

\subsection{System description and attack model}\label{sec2.1}
Let the discrete-time cyber-physical system of interest be modeled by
\be
x_{k+1} = \left\{ \ba{ll} 
f^0_k ( x_k ) + w_k,                   & \mbox{under no attack}       \vspace{2mm} \\
f^1_k ( x_k, a_k ) + w_k, &       \mbox{under attack}                                  
\ea \right.
\label{sys1}
\ee
where: 
$k$ is the time index; 
$x_k \in \mathbb{R}^n$ is the state vector to be estimated; 
$a_k \in \mathbb{R}^m$, called attack vector, is an unknown input affecting the system only when it is under attack;
$f_k^0(\cdot)$ and $f_k^1(\cdot,\cdot)$ are known state transition functions that describe the system evolution in the \textit{no attack} and, respectively, \textit{attack} cases;
$w_k$ is a random process disturbance also affecting the system.
For monitoring purposes, the state of the above system is observed through the measurement model
\be
y_k = \left\{ \ba{ll} 
h^0_k ( x_k ) + v_k,                   & \mbox{under no attack}       \vspace{2mm} \\
h^1_k ( x_k, a_k ) + v_k, &       \mbox{under attack}                                  
\ea \right.
\label{sys2}
\ee
where: 
$h_k^0(\cdot)$ and $h_k^1(\cdot,\cdot)$ are known measurement functions that refer to the \textit{no attack} and, respectively, \textit{attack} cases;	
$v_k$ is a random measurement noise.
It is assumed that the measurement $y_k$ is actually delivered  to the system monitor 
with probability $p_d \in (0,1]$, where the non-unit probability might be due to a number of reasons (e.g. temporary denial of service, packet loss, sensor inability to detect or sense the system, etc.).
The attack modeled in (\ref{sys1})-(\ref{sys2}) via the attack vector $a_k$ is usually referred to as 
\textit{signal} attack.
While for ease of presentation only the case of a single attack model is taken into account, multiple attack models
\cite{IFAC}
could be accommodated in the considered framework by letting (1)-(2) depend on a discrete variable, say $\nu_k$, which specifies the particular attack model and has to be estimated together with $a_k$.
Besides the system-originated measurement $y_k$ in (\ref{sys2}), 
it is assumed that the system monitor might receive \textit{fake} measurements from some cyber-attacker.
In this respect, the following two cases will be considered.
\begin{enumerate}
	\item \textit{Packet substitution} - With some probability $p_f \in [0,1)$, the attacker replaces the system-originated measurement $y_k$ with a fake one $y_k^f$.
	\item \textit{Extra packet injection} - The attacker sends to the monitor one or multiple fake measurements indistinguishable from the system-originated one.
\end{enumerate}
\begin{figure}[h!]
	\begin{center}
		\includegraphics[width=.47\columnwidth]{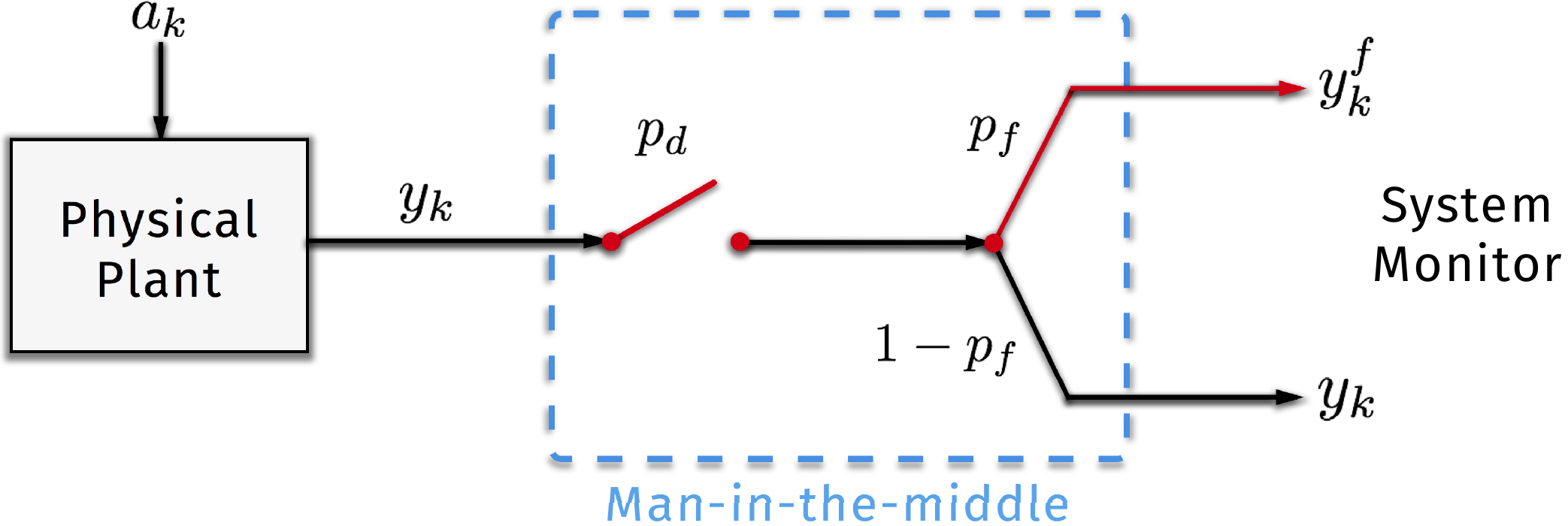}
		\caption{\textit{Packet substitution} attack.}
		\label{fig:ps}
	\end{center}
\end{figure} 
\begin{figure}[h!]
	\begin{center}
		\includegraphics[width=.53\columnwidth]{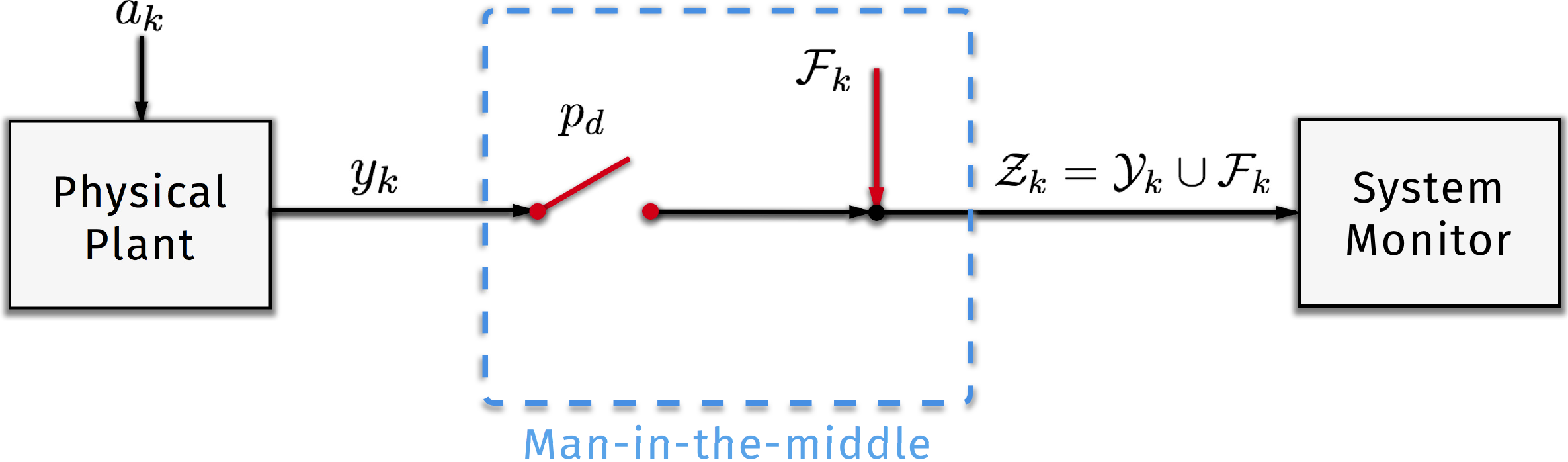}
		\caption{\textit{Extra packet injection} attack.}
		\label{fig:epi}
	\end{center}
\end{figure} 
For the subsequent developments, it is convenient to introduce the \textit{attack set} at time $k$, $\mathcal{A}_k$, which is either equal to the empty set if the system is not under 
signal attack at time $k$ or to the singleton $\{ a_k \}$ otherwise, i.e.
$$
\mathcal{A}_k ~=~ \left\{ \ba{cl} \emptyset, & \mbox{if the system is not under signal attack} \\
\{ a_k \}, & \mbox{otherwise}. \ea \right.
$$
It is also convenient to define the \textit{measurement set} at time $k$, $\mathcal{Z}_k$.
For the \textit{packet substitution} attack (Fig. \ref{fig:ps}):
\be
\mathcal{Z}_k ~=~ \left\{ \ba{cl} \emptyset, & \mbox{with probability } 1-p_d \vspace{1mm} \\
\{ y_k \},     & \mbox{with probability } p_d (1-p_f) \vspace{1mm} \\
\{ y_k^f \}, & \mbox{with probability } p_d \, p_f 
\ea \right.
\label{ps-attack}
\ee
where $y_k$ is given by (\ref{sys2}) and $y_k^f$ is a fake measurement provided by the attacker in place of $y_k$.
Conversely, for the \textit{extra packet injection} attack (Fig. \ref{fig:epi}) the definition (3) is replaced by
\be
\mathcal{Z}_k ~=~ \mathcal{Y}_k \cup \mathcal{F}_k
\label{epi-attack}
\ee
where
\be
\mathcal{Y}_k ~=~ \left\{ \ba{cl} \emptyset, & \mbox{with probability $1- p_d$} \\
\{ y_k \}, & \mbox{with probability $p_d$} \ea \right.
\label{meas1}
\ee
is the set of system-originated measurements and $\mathcal{F}_k$ the finite set of fake measurements.

The aim of this paper is to address the problem of joint attack detection and state estimation, which amounts to jointly estimating, at each time $k$, the state $x_k$ and signal attack set 
$\mathcal{A}_k$
given the set of measurements $\mathcal{Z}^k \defi \cup_{i=1}^k \mathcal{Z}_i$ up to time $k$.

\subsection{Joint input and state estimation}\label{sec2.2}

In this section, the main ideas of the Bayesian approach to \textit{Joint Input and State Estimation} (JISE) \cite{Fang2013} are summarized.
Consider a system affected by an unknown input $a_k$
\be
\left\{
\ba{rcl}
x_{k+1} &=& f ( x_k, a_k ) + w_k \\
y_k &=& h ( x_k, a_k ) + v_k 
\ea
\right.
\label{system_JISE}
\ee
In JISE \cite{Fang2013,Gillijns2007,Gillijns2007b} it is customary to distinguish the case in which there is a direct feedthrough 
of the unknown input $a_k$ to the output $y_k$ from the case of no direct feedthrough.
%Notice that, for the sake of simplicity, we only address the case of an invertible direct feedthrough 
%of the unknown input $a_k$ to the output $y_k$, \cite{Fang2013,Gillijns2007} which amounts to assuming 
%that the function $h ( x, a )$ is injective with respect to $a$  for any $x$. However, the following considerations 
%can be readily extended also to the case of no direct feedthrough. \cite{Fang2013,Gillijns2007b}
\vspace{.3cm}

\textit{Direct feedthrough:} Suppose that there is an invertible direct feedthrough\cite{Fang2013,Gillijns2007}
of the unknown input $a_k$ to the output $y_k$,  which amounts to assuming 
that the function $h ( x, a )$ is injective with respect to $a$  for any $x$. In this case,
the Bayesian approach is based on the recursive computation of the joint PDF $p(a_k,x_k|y^k)$ of the unknown input $a_k$ and state $x_k$ conditioned on all the information available up to the current time.
Given the conditional PDF, optimal estimates of $a_k$ and $x_k$ can be computed according to any given criterion, the most typical ones being Maximum A-posteriori Probability (MAP)  and
Minimum Mean Square Error (MMSE). The joint conditional PDF can be computed by means of a two-step procedure of correction and prediction.
Suppose that at time $k-1$, the predicted posterior $p(a_k , x_{k} |y^{k-1})$ has been computed. 
Then, at time $k$, when the new measurement $y_k$ is collected, in the correction step the new conditional PDF $p(a_k,x_k|y^k)$ can be obtained by means of the Bayes rule
\begin{equation}\label{eq:bayes}
	p(a_k,x_k|y^k) = \frac{p(y_k|a_k,x_k) \, p(a_k , x_{k}|y^{k-1}) }{p(y_k|y^{k-1})} 
\end{equation} 
Conversely, the prediction step concerns the propagation of the conditional PDF from time $k$ to time $k+1$. In the literature on unknown input estimation, it is usually supposed that
the values $a_k$ and $x_k$ of unknown input and, respectively, state at time $k$ do not provide any information on the value $a_{k+1}$ taken by the unknown input at time $k+1$.
Accordingly, $p(a_{k+1} , x_{k+1}|y^{k})$ takes the form
\begin{equation}\label{eq:ax}
	p(a_{k+1} , x_{k+1}|y^{k}) = p(x_{k+1}|y^{k}) \, p(a_{k+1}) 
\end{equation}
where the conditional PDF $p(x_{k+1}|y^{k})$ is computed via the Chapman-Kolmogorov equation
\begin{eqnarray} \label{eq:chapman}
	p(x_{k+1}|y^{k}) = \iint p(x_{k+1}|a_{k},x_{k}) \, p(a_{k},x_{k}|y^{k}) \, d a_{k} dx_{k} .
\end{eqnarray}
%and $p(a_k)$ is a PDF summarizing the prior knowledge on the input $a_k$. 
With this respect,
when no information on the unknown input $a_{k+1}$ is supposed to be available, it is customary \cite{Fang2013}
to resort to the so-called \textit{principle of indifference} and take 
$p(a_{k+1})$ as an uninformative (flat) prior.  It is easy to check that, in this case, the conditional PDF $p(a_k,x_k|y^k)$ resulting from the correction step can be rewritten as
\begin{equation}\label{eq:bayes2}
	p(a_k,x_k|y^k) = \frac{p(y_k|a_k,x_k) \, p(x_{k}|y^{k-1}) }{\int \int p(y_k|a,x) \, p(x|y^{k-1}) \, d x \, d a} 
\end{equation}
Then, maximization of (\ref{eq:bayes2}) with respect to $x_k$ and $a_k$ provides a  MAP estimate of $x_k$ and a Maximum Likelihood (ML) estimate of the unknown input $a_k$. This is the approach followed by Fang et al. \cite{Fang2013} that allows to generalize the traditional techniques for linear systems \cite{Gillijns2007,Gillijns2007b} to general nonlinear systems (see Theorems 1 and 2 in the work of Fang et al. \cite{Fang2013}).
\vspace{.3cm}

\textit{No direct feedthrough:}  Suppose that there is no direct feedthrough\cite{Fang2013,Gillijns2007b}
of the unknown input $a_k$ to the output $y_k$ so that $y_k= h(x_k) + v_k$. In this case, the
unknown input must be estimated with one step delay, since $y_{k+1}$ is the first measurement containing information on $a_k$.
Hence, the Bayesian approach is based on the recursive computation of the joint PDF $p(a_{k-1},x_k|y^k)$ of the unknown input $a_{k-1}$ and state $x_k$ conditioned on all the information available up to time $k$.
%Given the conditional PDF, optimal estimates of $a_{k-1}$ and $x_k$ can be computed according to any given criterion, the most typical ones being Maximum A-posteriori Probability (MAP)  and
%Minimum Mean Square Error (MMSE). The joint conditional PDF can be computed by means of a two-step procedure of correction and prediction.
Suppose that at time $k-1$, the predicted posterior $p(a_{k-1} , x_{k} |y^{k-1})$ has been computed. 
Then, at time $k$, when the new measurement $y_k$ is collected, in the correction step the new conditional PDF $p(a_{k-1},x_k|y^k)$ can be obtained by means of the Bayes rule
\begin{equation}
	p(a_{k-1},x_k|y^k) = \frac{p(y_k|x_k) \, p(a_{k-1}, x_{k}|y^{k-1}) }{p(y_k|y^{k-1})} 
\end{equation} 
%Conversely, the prediction step concerns the propagation of the conditional PDF from time $k$ to time $k+1$. In the literature on unknown input estimation, it is usually supposed that
%the values $a_k$ and $x_k$ of unknown input and, respectively, state at time $k$ do not provide any information on the value $a_{k+1}$ taken by the unknown input at time $k+1$.
%Accordingly, 
while in the prediction step, $p(a_{k} , x_{k+1}|y^{k})$ takes the form
\begin{equation}\label{eq:ax:no}
	p(a_{k} , x_{k+1}|y^{k}) = p(x_{k+1}|a_k , y^{k}) \, p(a_{k}) 
\end{equation}
where the conditional PDF $p(x_{k+1}|a_k,y^{k})$ is computed via the Chapman-Kolmogorov equation
\be
	p(x_{k+1}|a_k,y^{k}) = \int p(x_{k+1}|a_{k},x_{k}) \, p(x_{k}|y^{k}) \, dx_{k} .
\ee
%and $p(a_k)$ is a PDF summarizing the prior knowledge on the input $a_k$. 
When no information on the unknown input $a_{k}$ is supposed to be available so that
$p(a_{k})$ as an uninformative (flat) prior, the conditional PDF $p(a_{k-1},x_k|y^k)$ resulting from the correction step can be rewritten as
\begin{equation}
	p(a_{k-1},x_k|y^k) = \frac{p(y_k|x_k) \, p(x_{k}|a_{k-1} , y^{k-1}) }{\int \int p(y_k|x) \, p(x|a,y^{k-1}) \, d x \, d a} 
\end{equation}

\subsection{Random set estimation}\label{sec2.3}
An RFS (\textit{Random Finite Set}) $\mathcal{X}$ over $\mathbb{X}$ is a random variable taking values
in $\mathscr{F}(\mathbb{X})$, the collection of all finite subsets of  ${\mathbb{X}}$. 
The mathematical background needed for Bayesian random set estimation can be found in Mahler's book \cite{Mahler2007}; here, the basic concepts needed for the subsequent developments are briefly reviewed.
From a probabilistic viewpoint, an RFS $\mathcal{X}$ is completely characterized by its \textit{set density} $f(\mathcal{X})$, also called FISST (\textit{FInite Set STatistics}) probability density.
In fact, given $f(\mathcal{X})$, the cardinality \textit{probability mass function} $\rho(n)$ that $\mathcal{X}$ have $n \geq 0$ elements and the joint PDFs
$f \left( x_1, x_2, \dots, x_n | n \right)$ over $\mathbb{X}^n$ given that $\mathcal{X}$ have $n$ elements, are obtained as follows:
%While $\mathcal{F}({\mathbb{X}})$
%does not inherit the usual Euclidean notion of probability density from ${%
%\mathbb{X}}$, a measure-theoretic notion of probability density on $\mathcal{%
%F}({\mathbb{X}})$ is available \cite{doucet}. However, we adopt the \emph{%
%Finite Set Statistic} (FISST) notion of density since it is convenient and
%by-passes measure theoretic constructs \cite{Goodmanetal,mahler}.
%Hereafter, the basic concepts of FISST needed for the subsequent
%developments will be briefly reviewed.
$$
\ba{rcl}
\rho(n) & = & \dfrac{1}{n!} \, \displaystyle{\int_{\mathbb{X}^n}} f(\{x_1,\dots,x_n\}) \, dx_1 \cdots dx_n \vspace{1mm} \\
\hspace{-.2cm} f \left( x_1, x_2, \dots, x_n | n \right) & = & \dfrac{1}{n! \, \rho(n)} ~f(\{x_1,\dots,x_n\}) .
\ea
$$
In order to measure probability over subsets of $\mathbb{X}$ or compute expectations of random set variables, Mahler \cite{Mahler2007} introduced the notion of \textit{set integral}
for a generic real-valued function $g(\mathcal{X})$ of an RFS $\mathcal{X}$ as
\be
\int g(\mathcal{X}) \, \delta \mathcal{X} = g(\emptyset) + \sum_{n=1}^{\infty} \frac{1}{n!} \int g(\{x_1,\dots,x_n\}) \, dx_1 \cdots dx_n
\label{eq:integral}
\ee
%Two specific types of RFSs, i.e. Bernoulli and Poisson RFSs, will be considered in this work.
In particular, in this work we will consider the Bernoulli RFS, i.e.
a random set which can be either empty or, with some probability $r \in [0,1]$, a singleton $\{ x \}$ whose element is distributed over $\mathbb{X}$ according to the PDF $p(x)$. Accordingly, its set density is defined as follows:
\begin{equation}
	f(\mathcal{X})=\begin{cases}
		1-r, & \text{if $\mathcal{X} = \emptyset $}\\
		r \cdot p(x), & \text{if $\mathcal{X} = \{ x \}$}
	\end{cases}
\end{equation}
Please notice that the above equation as well as all subsequent definitions of probability distributions involving a Bernoulli set argument have two branches on the right-hand-side depending on whether the Bernoulli argument is empty or a singleton.

%\subsection*{Poisson RFS}
%A Poisson RFS is a random finite set with Poisson-distributed cardinality, i.e.
%\begin{equation}
%	p(n) = \frac{e^{-\xi}\xi^n}{n!}, \,\,\, n = 0,1,2,\dots
%	\label{poisson_cardinality}
%	\end{equation}
%and elements independently distributed over $\mathbb{X}$ according to a given spatial density $p(\cdot)$.
%Accordingly, its set density is defined as follows:
%	\begin{equation}
%	f(\mathcal{X}) = e^{-\xi}  \prod_{x \in \mathcal{X}} \xi \, p(x).
%	\label{poisson_pdf}
%	\end{equation}	

\section{Bayesian Random Set Filter for Joint Attack Detection and State Estimation -- the direct feedthrough case}\label{sec3}

Let us suppose that, when the attack input is present, there is a direct  feedthrough  from the attack $a_k$ to the output $y_k$. More specifically, 
in accordance with the considerations of Section~\ref{sec2.2}, it is assumed that, when the attack input is present, the mapping from $a_k$ to $y_k$ is full rank, i.e. invertible.
Let the attack input at time $k$ be modeled as a Bernoulli random set $\A_k \in \mathscr{B}(\mathbb{A})$, where $\mathscr{B}(\mathbb{A}) = \emptyset \, \cup \, \mathscr{S}(\mathbb{A})$ is a set of all finite subsets of the attack space $\mathbb{A} \subseteq \mathbb{R}^m$, and $\mathscr{S}(\mathbb{A})$ denotes the set of all singletons (i.e., sets with cardinality 1) $\{ a \}$ such that $a \in \mathbb{A}$.
Further, let $\mathbb{X} \subseteq \mathbb{R}^n$ denote the Euclidean space for the system state vector, 
then we can define the \textit{Hybrid Bernoulli Random Set} (HBRS) $(\A,x)$ as a new state variable which incorporates the Bernoulli attack random set $\A$ and the random state vector $x$, taking values in the hybrid space $\mathscr{B}(\mathbb{A}) \times \mathbb{X}$.
A HBRS is fully specified by the (signal attack) probability $r$ of $\mathcal{A}$ being a singleton, the PDF $p^0(x)$ defined on the state space $\mathbb{X}$, and the joint PDF $p^1(a,x)$ defined on the joint attack input-state space $\mathbb{A} \times \mathbb{X}$, i.e. 
\be
p(\A,x) = \left\{ \ba{ll} 
(1 - r) \, p^0(x),                   & \mbox{if } \A = \emptyset      \vspace{2mm} \\
r \cdot p^1(a,x),                   & \mbox{if } \A = \{a\}                           
\ea  \right..   
\label{eq:joint_bernoulli}
\ee
Moreover, since integration over $\mathscr{B}(\mathbb{A}) \times \mathbb{X}$ takes the form
\begin{equation}
	\int_{\mathscr{B}(\mathbb{A}) \times \mathbb{X}} p(\A,x) \delta \A \, dx = \int p(\emptyset,x) \, dx + \iint p(\{a\},x) \, da \, dx
	\label{joint_integral}
\end{equation}
where the set integration with respect to $\A$ is defined according to (\ref{eq:integral}) while the integration with respect to $x$ is an ordinary one,
it is easy to see that $p(\A,x)$ integrates to one by substituting (\ref{eq:joint_bernoulli}) in (\ref{joint_integral}), and noting that $p^0(x)$ and $p^1(a,x)$ are conventional probability density functions on $\mathbb{X}$ and $\mathbb{A} \times \mathbb{X}$, respectively.
This, in turn, guarantees that (\ref{eq:joint_bernoulli}) is a FISST probability density for the HBRS $(\A,x)$. %, which will be
%referred to as \textit{hybrid Bernoulli density} throughout the rest of the paper. 
The notion of \textit{attack existence}, embodied by parameter $r$ in \eqref{eq:joint_bernoulli}, is introduced so as to detect the presence (existence) of a signal attack and hence initiate its estimation. Thanks to this concept, as shown later on, the probability of attack existence is directly computed by the filter.

In this paper the attack input is modeled as a Bernoulli random set (BRS) to account for the fact that the attack can switch (from \textit{off} to \textit{on} or viceversa)
at any time with no prior knowledge on the attack onset/termination from the system monitor side.
The switching nature of the attack could be tackled in different ways, e.g. with multiple models (one for the attack and another for the no-attack cases), but the random set approach undertaken in this work turns out to be advantageous also to include other type of attacks, specifically packet substitution and extra packet injection to be considered in the next subsection.
\vspace{.3cm}
	
\subsection{Measurement models and correction}\label{sec3.1}

\subsubsection{Packet substitution}\label{sec3.1.1}

%\vspace{.3cm}
Let us consider the \textit{packet substitution} attack model introduced in Section 2.1 and denote by $\lambda(\Z_k|\A_{k},x_{k})$ the likelihood function of the measurement set defined in (\ref{ps-attack}), which has obviously two possible forms, $\A_{k}$ being a Bernoulli random set.
In particular, for $\A_{k} = \emptyset$:
	\be
	\lambda(\Z_k|\emptyset,x_{k}) = \left\{ \ba{ll} 
%	\hspace{-.2cm}	
	1 - p_d,                 & 
%	\hspace{-.2cm} 
	\mbox{if } \Z_{k} = \emptyset      \vspace{2mm} \\
%	\hspace{-.2cm}	
	p_d \big[(1 - p_f) \, \ell(y_{k}|x_{k}) + p_f \, \kappa(y_{k})\big],                   & 
%	\hspace{-.2cm} 
	\mbox{if } \Z_{k} = \{y_{k}\}      
	\ea \right.
	\label{eq:likelihood_A0}
	\ee 
where $\{y_k\}$ denotes the singleton whose element represents a delivered measurement, 
%which can either be $y_k$ or $y_k^f$, 
i.e. $\lambda(\{y_k\}|\A_{k},x_{k})$ is the likelihood that a single measurement $y_k$ will be collected.
Furthermore, $\ell(y_{k}|x_{k})$ is the standard likelihood function of the system-generated measurement $y_k$ when no signal attack is present, 
%i.e. the probability density that, at time $k$, the system with state $x_k$ produces a measurement $y_k$,
whereas $\kappa(\cdot)$ is a PDF modeling the fake measurement $y_k^f$, assumed to be independent of the system state.
Conversely, for $\A_{k} = \{a_k\}$:
\begin{eqnarray}\label{eq:likelihood_A1} 
	\lambda(\Z_k|\{ a_{k} \},x_{k}) = \left\{ \ba{ll} 
	1 - p_d,                  & \mbox{if } \Z_{k} = \emptyset      \vspace{2mm}   \\
	p_d \big[(1 - p_f) \, \ell(y_{k}|a_k,x_{k}) + p_f \, \kappa(y_{k})\big],                   & \mbox{if } \Z_{k} = \{y_{k}\}     
	%\vspace{2mm} \\
	%p_d \, p_f \, g(y^f_{k}),                   & \mbox{if } \Z_{k} = \{y^f_{k}\}                          
	\ea \right.  
\end{eqnarray} 
where $\ell(y_{k}|a_k,x_{k})$ denotes the conventional likelihood of measurement $y_k$, due to the system under attack $a_k$ in state $x_k$. 
Notice that, by using the definition of set integral (\ref{eq:integral}), it is easy to check that both forms 
(\ref{eq:likelihood_A0}) and (\ref{eq:likelihood_A1}) of the likelihood function $\lambda(\Z_k|\A_{k},x_{k})$ integrate to one.
Using the aforementioned measurement model, it is possible to derive the exact correction equations of the Bayesian random set filter for joint attack detection and state estimation, in case of substitution attack.
\vspace{.3 cm}

\begin{theorem} (\textit{Correction under packet substitution attack})
	Suppose that the prior density 
	%	$p_{k|k-1}(\A_{k},x_{k}|\Z^{k-1})$ 
	at time $k$ is \textit{hybrid Bernoulli} of the form
	\be
		p(\A_{k},x_{k}|\Z^{k-1}) = \left\{ \ba{ll} 
		(1 - r_{k|k-1}) \, p^0_{k|k-1}(x_{k}),                   & \mbox{if } \A_{k} = \emptyset      \vspace{2mm} \\
		r_{k|k-1} \cdot p^1_{k|k-1}(a_{k},x_{k}),                   & \mbox{if } \A_{k} = \{a_{k}\}                                  
		\ea \right..
	\label{eq:prediction_0}
	\ee
	Then, given the measurement random set $\Z_k$ defined in (\ref{ps-attack}), 
	also the posterior density at time $k$ turns out to be \textit{hybrid Bernoulli} of the form
	\be
	p(\A_k,x_k|\Z^k) = \left\{ \ba{ll} 
	(1 - r_{k|k}) \, p^0_{k|k}(x_k),                   & \mbox{if } \A_k = \emptyset      \vspace{2mm} \\
	r_{k|k} \cdot p^1_{k|k}(a_k,x_k),                  & \mbox{if } \A_k = \{a_k\}                                  
	\ea \right.
	\label{eq:correction_final}
	\ee
	completely specified by the triplet
		\begin{equation*}
			\ba{l}
%			\hspace{-.2cm}
			\big( r_{k|k}, p^0_{k|k}(x_k), p^1_{k|k}(a_k,x_k) \big) = 
			\big( r_{k|k-1}, p^0_{k|k-1}(x_k), p^1_{k|k-1}(a_k,x_k) \big)
			\ea
		\end{equation*}
	if $\Z_k = \emptyset$
	or, if $\Z_k = \{y_k\}$, by:
	\begin{eqnarray} 
		r_{k|k} &&
%		\hspace{-.6cm} 
		= \frac{(1 - p_f) \, \Psi_1 + p_f \kappa(y_k)}{(1 - p_f) (\Psi_0 - r_{k|k-1} \Psi) + p_f \kappa(y_k)} \, r_{k|k-1}  \label{eq:correction_r} \\
		p^0_{k|k}(x_{k}) && 
%		\hspace{-.6cm} 
		= \frac{(1 - p_f) \, \ell(y_k|x_k) + p_f \kappa(y_k)}{(1 - p_f) \, \Psi_0 + p_f \kappa(y_k)} \, p^0_{k|k-1}(x_k)  \label{eq:correction_0} \\
		p^1_{k|k}(a_{k},x_{k}) && 
%		\hspace{-.6cm} 
		= \frac{(1 - p_f) \, \ell(y_k|a_k,x_k) + p_f \kappa(y_k)}{(1 - p_f) \, \Psi_1 + p_f \kappa(y_k)} \, p^1_{k|k-1}(a_k,x_k) \label{eq:correction_1}
	\end{eqnarray} 
	where
	\begin{eqnarray}
		\Psi_0 & \defi &\int \ell(y_{k}|x_{k}) \, p^0_{k|k-1}(x_k) \, \mbox{d} x_{k} \label{psi_0}  \\
		\Psi_1 & \defi & \iint \ell(y_{k}|a_k,x_{k}) \, p^1_{k|k-1}(a_k,x_k) \, \mbox{d} a_{k} \mbox{d} x_{k} \label{psi_1} \\
		\Psi & \defi &\Psi_0 - \Psi_1.
	\end{eqnarray} 
\end{theorem}  	 	
%\hfill $\square$
%\vspace{.3cm}
%{\em Proof:} 

\textit{Proof:}
The correction equation of the Bayes random set filter for joint attack detection and state estimation
follows from a generalization of (\ref{eq:bayes}), which yields
\be
p(\A_k,x_k|\Z^k) = \frac{\lambda(\Z_k|\A_{k},x_{k}) \, p(\A_{k},x_{k}|\Z^{k-1})}{p(\Z_k|\Z^{k-1})}
\label{eq:correction_bayes}
\ee
where $\lambda(\Z_k|\A_{k},x_{k})$ is given by (\ref{eq:likelihood_A0}) and (\ref{eq:likelihood_A1}), while 
\begin{eqnarray} 
	p(\Z_k|\Z^{k-1}) && 
%	\hspace{-.6cm} 
	= \iint \lambda(\Z_k|\A_{k},x_{k})  	\label{eq:update_den}  \, p(\A_{k},x_{k}|\Z^{k-1}) \, \delta \A_{k} \mbox{d} x_{k} \nonumber \\
	&& 
%	\hspace{-1.6cm} 
	= \int \lambda(\Z_k|\emptyset,x_{k}) \, p(\emptyset,x_{k}|\Z^{k-1}) \, \mbox{d} x_{k} 
%	\nonumber \\
%	&& 
%	\hspace{-1.6cm} 
	+ \, \iint \lambda(\Z_k|\{a_{k}\},x_{k}) \, p(\{a_k\},x_{k}|\Z^{k-1}) \, \mbox{d} a_{k} \mbox{d} x_{k}  .
	\label{eq:update_den2}
\end{eqnarray} 
For the case $\Z_k = \emptyset$, the above reduces to
\be
p(\emptyset|\Z^{k-1}) = 1 - p_d
\label{eq:den0}
\ee
by substituting (\ref{eq:likelihood_A0})-(\ref{eq:likelihood_A1}) and (\ref{eq:prediction_0}) in (\ref{eq:update_den2}), and simply noting that $\int p^0_{k|k-1}(x_{k}) \mbox{d} x_{k} = 1$ and $\iint p^1_{k|k-1}(a_{k},x_{k}) \, \mbox{d} a_{k} \mbox{d} x_{k} = 1$.
The posterior probability of attack existence $r_{k|k}$ can be obtained from the posterior density (\ref{eq:correction_bayes}) with $\A_k = \emptyset$ via 
\be
r_{k|k} = 1 - \int p(\emptyset,x_k|\Z^k) \, \mbox{d} x_{k}
\label{p02r}
\ee
where - using (\ref{eq:likelihood_A0}), (\ref{eq:prediction_0}) and (\ref{eq:den0}) in (\ref{eq:correction_bayes}) - we have
\be
p(\emptyset,x_k|\Z^k) = (1 - r_{k|k-1}) \, p^0_{k|k-1}(x_{k})
\label{eq:corr0}.
\ee
Moreover, $p^0_{k|k}(x_k) = p(\emptyset,x_k|\Z^k) / (1 - r_{k|k})$, and the joint %spatial 
density for the system under attack can be easily derived from the posterior density with $\A_k = \{a_k\}$ by recalling that 
$p^1_{k|k}(a_{k},x_{k}) = p(\{ a_k \},x_k|\Z^k) / r_{k|k}$, where 
\be
p(\{ a_k \},x_k|\Z^k) = r_{k|k-1} \cdot p^1_{k|k-1}(a_k,x_k)
\label{eq:corr1}
\ee 
results from replacing (\ref{eq:likelihood_A1}), (\ref{eq:prediction_0}) and (\ref{eq:den0}) in (\ref{eq:correction_bayes}).
Notice that from the set integral definition (\ref{eq:integral}), and densities (\ref{eq:corr0})-(\ref{eq:corr1}), it holds that $\int p(\emptyset,x_k|\Z^k) \, \mbox{d} x_{k} + \iint p(\{ a_k \},x_k|\Z^k) \, \mbox{d} a_{k} \mbox{d} x_{k} = 1$. 
Hence, as stated, the Bayes correction (\ref{eq:correction_final}) provides a hybrid Bernoulli density.
Next, for the case $\Z_k = \{ y_k \}$, (\ref{eq:update_den2}) leads to
\be
p(\{y_k\}|\Z^{k-1}) = p_d \bigg[ (1 - p_f) (\Psi_1 - r_{k|k-1} \Psi)  + p_f \kappa(y_k) \bigg]
\ee
so that from (\ref{eq:correction_bayes}) one gets
\begin{eqnarray} 
	&& p(\emptyset,x_k|\Z^k) = \frac{\bigg[ (1 - p_f) \, \ell(y_k|x_k) + p_f \kappa(y_k) \bigg]}{(1 - p_f) (\Psi_1 - r_{k|k-1} \Psi) + p_f \kappa(y_k)} (1 - r_{k|k-1}) \, p^0_{k|k-1}(x_k)  
\end{eqnarray} 
which, in turn, is used to obtain (\ref{eq:correction_r}) through (\ref{p02r}).
Once $r_{k|k}$ is known, (\ref{eq:correction_0}) immediately follows as previously shown for the case $\Z_k = \emptyset$, while (\ref{eq:correction_1}) comes from dividing the posterior
\begin{eqnarray} 
	&&	p(\{ a_k \},x_k|\Z^k) = \frac{\bigg[ (1 - p_f) \, \ell(y_k|x_k) + p_f \kappa(y_k) \bigg]}{(1 - p_f) (\Psi_1 - r_{k|k-1} \Psi) + p_f \kappa(y_k)} r_{k|k-1} \, p^1_{k|k-1}(a_k,x_k)   
\end{eqnarray}
by $r_{k|k}$ in (\ref{eq:correction_r}).

% \qed

\subsubsection{Extra packet injection}\label{sec3.1.2}

A complete derivation of the correction step for the \textit{extra packet injection} model introduced in Section \ref{sec2.1} can be found in Forti et al. \cite{TAC}
We summarize below the main results, since they are the basis for the derivation of the Gaussian-mixture filter of Section 4.
First recall that, in this case, the measurement set $\Z_k$ is given by the union of the two independent random sets $\Y_k$ and $\F_k$.
Clearly, in view of  (\ref{meas1}), $\Y_k$ is a Bernoulli random set whose cardinality is either $0$ or $1$ depending on whether the system-originated measurement $y_k$ is delivered or not.
Conversely, it is supposed that no prior knowledge on the number of fake measurements, i.e. the cardinality of $\F_k$, is available.
Accordingly, $\rho(n)$ is taken as an uninformative distribution and, hence, the
FISST PDF of fake-only measurements turns out to be
\begin{equation}\label{eq:fake}
	\gamma (\F_k) \propto |\F_k|! \, \prod_{y_k \in \F_k} \kappa(y_k)
\end{equation}
where $\kappa(y_k)$ is a PDF describing the distribution of fake measurements on the measurement space $\mathbb Y$.
Clearly, if no prior knowledge on such a distribution can be assumed,  the same approach of Section 2.1 can be followed by taking $\kappa(y_k)$ as an uninformative (i.e. uniform) PDF over  $\mathbb Y$.
The following result holds.
\vspace{.3 cm}

\begin{theorem}
	(\textit{Correction under extra packet injection attack}, Forti et al. \cite{TAC})
	%	Assume that the prior density 
	%	at time $k$ is hybrid Bernoulli of the form (\ref{eq:prediction_-1}).
	%Then, given the measurement random set $\Z_k$ defined in (\ref{epi-attack}), 
	%also the posterior density at time $k$ turns out to be hybrid Bernoulli of the form (\ref{eq:correction_final}),  
	Suppose that the prior density 
	at time $k$ is \textit{hybrid Bernoulli} of the form
	\be \label{eq:prediction_-1}
		p(\A_{k},x_{k}|\Z^{k-1}) = \left\{ \ba{ll} 
		(1 - r_{k|k-1}) \, p^0_{k|k-1}(x_{k}),                   & \mbox{if } \A_{k} = \emptyset      \vspace{2mm} \\
		r_{k|k-1} \cdot p^1_{k|k-1}(a_{k},x_{k}),                   & \mbox{if } \A_{k} = \{a_{k}\}                                  
		\ea \right..
	\ee
	Then, given the measurement random set $\Z_k$ defined in (\ref{epi-attack}), 
	also the posterior density at time $k$ turns out to be \textit{hybrid Bernoulli} of the form
	\be
	p(\A_k,x_k|\Z^k) = \left\{ \ba{ll} 
	(1 - r_{k|k}) \, p^0_{k|k}(x_k),                   & \mbox{if } \A_k = \emptyset      \vspace{2mm} \\
	r_{k|k} \cdot p^1_{k|k}(a_k,x_k),                  & \mbox{if } \A_k = \{a_k\}                                  
	\ea \right.
	%\label{eq:correction_final}
	\ee
	completely specified by the triplet
	\begin{eqnarray}  
		r_{k|k} && = \frac{1 - p_d \, (1 -  \Gamma_1)}{1 - p_d [1 -  (\Gamma_0 - r_{k|k-1} \Gamma) ]} \, r_{k|k-1}  
		\label{p_existence3} 
		%r_{k|k} && \hspace{-.6cm} = \frac{1 - p_d+ p_d \Gamma_1}{1 - p_d [1 -  (\Gamma_0 - r_{k|k-1} \Gamma) ]} \, r_{k|k-1}  
		%		\label{p_existence3} 
		\\
		p^0_{k|k}(x_k) && = \frac{1 - p_d + p_d \displaystyle \sum_{y_k \in \Z_k} \frac{\ell(y_k|x_k)}{n \, \kappa(y_k)} }{1 - p_d \, (1 -  \Gamma_0)} \, p^0_{k|k-1}(x_k) 
		\label{p_03} 
		%		\\
		%		p^1_{k|k}(a_k,x_k) && \hspace{-.6cm} = \frac{1 - p_d \, \bigg[ 1 - \frac{1}{n} \displaystyle \sum_{y_k \in \Z_k} \frac{\ell(y_k|a_k,x_k)}{n \, \kappa(y_k)}\bigg]}{1 - p_d \, (1 - \frac{1}{n} \Gamma_1)} \, p^1_{k|k-1}(a_k,x_k)  
		%		\nonumber \\ \label{p_13}  
		\\
		p^1_{k|k}(a_k,x_k) && = \frac{1 - p_d + p_d \displaystyle \sum_{y_k \in \Z_k} \frac{\ell(y_k|a_k,x_k)}{n \, \kappa(y_k)} }{1 - p_d \, (1 -  \Gamma_1)} \, p^1_{k|k-1}(a_k,x_k)  
\label{p_13}  
	\end{eqnarray} 
	where
	%\hspace{-.4cm} 
	\begin{eqnarray} 
		\Gamma_0 & \defi & \sum_{y_k \in \Z_k} \frac{\int \ell(y_{k}|x_{k}) \, p^0_{k|k-1}(x_k) \, \mbox{d} x_{k}}{n \, \kappa(y_k)} \label{gamma_0}\\
		\Gamma_1 & \defi & \sum_{y_k \in \Z_k} \frac{\iint \ell(y_{k}|a_k,x_{k}) \, p^1_{k|k-1}(a_k,x_k) \, \mbox{d} a_{k} \mbox{d} x_{k}}{n \, \kappa(y_k)} 
		\label{gamma_1}
	\end{eqnarray}
	and $\Gamma \defi \Gamma_0 - \Gamma_1$. 	 	
	%\hfill $\square$
\end{theorem}
%\vspace{.3 cm}

\subsection{Dynamic model and prediction}\label{sec3.2}

Let us now focus on the prediction step of the Bayesian hybrid Bernoulli filter.
Concerning the propagation of the signal attack from time $k$ to time $k+1$, we consider the most general model for signal attacks where any value can be injected
and, accordingly, we model $a_{k+1}$ as a completely unknown input whose value does not depend on the values $a_k$ and $x_k$ of attack and, respectively, state at time $k$.
However, concerning the existence of the attack at time $k+1$, we introduce two parameters $p_s$ and $p_b$ to model the fact 
that the presence of an attack at time $k+1$ is more probable when an attack is already present at time $k$: $p_b$ denotes the probability that
an attack $a_{k+1}$ is launched to the system at time $k+1$ when the system is under normal operation at time $k$;
$p_s$ denotes the probability that an adversarial action affecting the system at time $k$ will endure to time $k+1$. 
Notice that the probabilities $p_b$ and $p_s$ have to be regarded as design parameters for the filter that can be tuned depending on the desired properties: 
the lower is $p_b$ the more cautious will be the filter in declaring the presence of an attack; the higher is $p_s$ the more cautious will be the filter in declaring that the attack has disappeared.
According to this model, the transition density $\pi(\A_{k+1}|\A_{k})$ of the attack BRS takes the form
\begin{eqnarray} 
	\pi(\A_{k+1}|\emptyset) &=& \left\{ \ba{ll} 
	1 - p_b,                   & \mbox{if } \A_{k+1} = \emptyset      \vspace{2mm} \\
	p_b \, p(a_{k+1}),                  & \mbox{if } \A_{k+1} = \{a_{k+1}\}                                  
	\ea \right. \nonumber
	\label{eq:markov_A0}
	\\	
	\pi(\A_{k+1}|\{a_k\}) &=& \left\{ \ba{ll} 
	1 - p_s,                  & \mbox{if } \A_{k+1} = \emptyset      \vspace{2mm} \\
	p_s \, p(a_{k+1}),                  & \mbox{if } \A_{k+1} = \{a_{k+1}\}                                  
	\ea \right. \nonumber
	\label{eq:markov_A1}
\end{eqnarray}
Like in Section 2.2, $p(a_{k+1})$ is the PDF summarizing the available knowledge on  $a_{k+1}$, which can be taken equal to an uninformative PDF 
(e.g., uniform over the attack space) when the attack vector is completely unknown. 

Then, the joint transition density of $(\A,x)$  at time $k+1$ takes the form
\be
\pi(\A_{k+1},x_{k+1}|\A_k,x_k) = \pi(x_{k+1}|\A_k,x_k) \, \pi(\A_{k+1}|\A_{k}) 
\label{markov_ax}
\ee
where, in accordance with (\ref{sys1}), we have
\be
\pi(x_{k+1}|\A_k,x_k) = \left\{ \ba{ll} 
\pi(x_{k+1}|x_k),                  & \mbox{if } \A_k = \emptyset      \vspace{2mm} \\
\pi(x_{k+1}|a_k,x_k),                & \mbox{if } \A_k = \{a_k\}                                  
\ea \right.
\label{eq:markov_x}
\ee
with  $\pi(x_{k+1}|x_k)$ and $\pi(x_{k+1}|a_k,x_k)$ known Markov transition PDFs. 

%%%%%%%%%%%%%%%%%%%%%%%%%%%%%%%%%%%%%%%%%%%%%%%%%%%%%%%%%%%%%%%%%%%%%%%%%%%%%%%%%%%%%%%%%%%%%%%
%%%%%%%%%%%%%%%%%%%%%%%%%%%%%%%%%%%%%%%%%%%%%%%%%%%%%%%%%%%%%%%%%%%%%%%%%%%%%%%%%%%%%%%%%%%%%%%

Under the above assumptions, Forti et al. \cite{TAC} obtained an exact recursion for the prior density.
\vspace{.3 cm}

\begin{theorem}
	(Forti et al. \cite{TAC})
	Given the posterior hybrid Bernoulli density 
	$p(\A_{k},x_{k}|\Z^{k})$ 
	at time $k$ of the form (\ref{eq:correction_final}),
	fully characterized by the triplet $\big( r_{k|k}, p^0_{k|k}(x_k), p^1_{k|k}(a_k,x_k) \big)$,
	also the predicted density turns out to be hybrid Bernoulli of the form 
	\begin{eqnarray} \label{eq:prediction_final}	
		&& p(\A_{k+1},x_{k+1}|\Z^{k}) =  \left\{ \ba{ll} 
		(1-r_{k+1|k}) \, p^0_{k+1|k}(x_{k+1}),                   &   \mbox{if } \A_{k+1} = \emptyset      \vspace{2mm}   \\
		r_{k+1|k} \cdot p^1_{k+1|k}(a_{k+1},x_{k+1}),                   &   \mbox{if } \A_{k+1} = \{a_{k+1}\}                                 
		\ea \right.  
	\end{eqnarray} 
	with
	\begin{eqnarray} 
		r_{k+1|k} = && (1 - r_{k|k}) \, p_b + r_{k|k} \, p_s  \label{p_existence1} \\
		p^0_{k+1|k}(x_{k+1}) =&&  \frac{(1 - r_{k|k}) (1 - p_b) \, p_{k+1|k}(x_{k+1}|\emptyset)}{1 - r_{k+1|k}} 
		+ \frac{r_{k|k} (1 - p_s) \, p_{k+1|k}(x_{k+1}|\{a_{k}\})}{1 - r_{k+1|k}}  \label{p_01} \\
		p^1_{k+1|k}(a_{k+1},x_{k+1}) = &&  \frac{(1 - r_{k|k}) \, p_b \, p_{k+1|k}(x_{k+1}|\emptyset) \, p(a_{k+1})}{r_{k+1|k}}  + \frac{r_{k|k} \, p_s \, p_{k+1|k}(x_{k+1}|\{a_{k}\}) \, p(a_{k+1})}{r_{k+1|k}} \label{p_11}
	\end{eqnarray} 
	where  
	\begin{eqnarray} 
		p_{k+1|k}(x_{k+1}|\emptyset) =&& \int \pi(x_{k+1}|x_{k}) \, p^0_{k|k}(x_{k}) \, \mbox{d} x_{k}  \label{def1} \\
		p_{k+1|k}(x_{k+1}|\{a_{k}\}) =&& \iint \pi(x_{k+1}|a_{k},x_{k}) \, p^1_{k|k}(a_{k},x_{k}) \,  \mbox{d} a_{k} \mbox{d} x_{k}  .
\label{def2} 	  
	\end{eqnarray} 	
	\normalsize
	%	\hfill $\square$
\end{theorem} 

Notice that, if $p_b = 0$, $p_s = 1$ and $r_{k|k} = 1$, it follows that $r_{k+1|k} = 1$ %, $p^0_{k+1|k}(x_{k+1}) = 0$, 
and $p^1_{k+1|k}(a_{k+1},x_{k+1}) = p_{k+1|k}(x_{k+1}|\{a_{k}\}) \, p(a_{k+1})$.
Hence, in this case, we recover the standard Chapman--Kolmogorov equation  (\ref{eq:chapman}) for the system under attack.
\vspace{.3 cm}

\begin{remark}
	Given the conditional density $p( \mathcal A_k, x_k | \Z^k)$, characterized by the triplet
	$\left( r_{k|k}, p^0_{k|k}(\cdot), p^1_{k|k}(\cdot,\cdot) \right)$, the joint attack detection and state estimation problem can be solved as follows. 
	First of all, we perform attack detection using $r_{k|k}$ from the available current hybrid Bernoulli density
	$p(\A_k,x_k|\Z^k)$. 
	By using a MAP decision rule, given $\Z_k$, the detector will assign $\hat{\A}_k \neq \emptyset$ (the system is under attack) if and only if 
	$\text{Prob}(\A_k \neq \emptyset | \Z^k) > \text{Prob}(\A_k = \emptyset | \Z^k)$, 
	i.e. if and only if $r_{k|k} > 1/2$.
	Then, if the signal attack has been detected, one can maximize $p( \mathcal A_k, x_k | \Z^k)$ with respect to $x_k$ and $a_k$.
	In this way it is possible to obtain a MAP estimate of $x_k$ and an ML estimate of the unknown attack input $a_k$. \\
\end{remark}
\vspace{.3 cm}

\begin{remark}
The Bayesian formulation of this section has allowed to generalize the standard \textit{joint input and state} filtering process to take into account
several practically relevant issues like the switching nature of the attack input, the injection of fake measurements or replacement of  system-originated by fake measurements, and the possible lack of system-originated measurements.
Please notice that all such phenomena are not contemplated in the standard filtering process.
%As also remarked in the previous point, the resulting equations can be (approximately) solved by a Gaussian-mixture approach but could also be solved by particle filtering though  the latter approach is not considered in this paper.
%The Gaussian-mixture implementation of section 4 actually reveals a connection between the proposed HBRS filter and the Kalman filter (KF) in that the former uses multiple KFs
%(or EKFs/UKFs) to propagate in time means and covariances of the various components of the Gaussian mixture
%(see eqns. (77)-(80), (87)-(95), (121)-(122) and (124)-(125)).
\end{remark}
\vspace{.3 cm}

\begin{remark}
 The HBRS Bayesian filtering recursions derived in this section are rarely solvable in explicit form but, as it will be shown in the next section, this is possible in the linear-Gaussian case.	
 In such a case, in fact, the propagated PDFs $p_{k|k}^0(\cdot)$ and $p_{k|k}^1(\cdot,\cdot)$ turn out to be Gaussian mixtures at any time $k$, even if with a number of Gaussian components growing with time and hence to be reduced via suitable pruning \& merging procedures.
\end{remark}
\vspace{.3 cm}

\begin{remark}
It is clear from the previous derivations that the defense method against signal attacks is embedded in the proposed hybrid Bernoulli filter and can be coordinated with any of the defense methods against the two considered data attacks, either packet substitution or extra packet injection.
In fact, it suffices to perform the correction step of the HBF according to either Theorem 1 or Theorem 2 while the prediction step is clearly unaffected by the 
choice of the data attack model.
Please notice that packet substitution and extra packet injection attacks are clearly alternative and that the HBF can switch from counteracting one or the other at any time, just by choosing the appropriate correction step, depending on whether the system monitor receives a single or multiple data packets during the sampling interval. The above described strategy could, therefore,
provide a sensible way to coordinate the defense methods against packet substitution and extra packet injection cyber-attacks.
\end{remark}

\section{Bayesian Random Set Filter for Joint Attack Detection and State Estimation -- the no direct feedthrough case}\label{sec4}

Suppose now that, even when the attack input is present, there is no direct  feedthrough from the attack $a_k$ to the output $y_k$, so that
the measurement model is
\begin{equation}
y _k = h(x_k) + v_k
\end{equation}
irrespectively of the presence of the attack. In this case, clearly, the attack set $\A_k$ must be estimated with one step delay,
since $\Z_{k+1}$ is the first measurement set containing information on $\A_k$.
In the following sections, a detailed derivation of the correction and prediction steps of the Bayes recursion in the case of no direct feedthrough is provided.

\subsection{Measurement models and correction}\label{sec4.1}

In the case of packet substitution with no direct feedthrough, the likelihood function $\lambda(\Z_k|x_{k})$ takes the following form:
	\be
	\lambda(\Z_k|x_{k}) = \left\{ \ba{ll} 
	1 - p_d,                 & 
	\mbox{if } \Z_{k} = \emptyset      \vspace{2mm} \\
	p_d \big[(1 - p_f) \, \ell(y_{k}|x_{k}) + p_f \, \kappa(y_{k})\big],                   & 
	\mbox{if } \Z_{k} = \{y_{k}\}      
	\ea \right.
	\ee 
where $\ell(y_{k}|x_{k})$ is the standard likelihood function of the system-generated measurement $y_k$.
It is easy to check that the likelihood function $\lambda(\Z_k|x_{k})$ integrates to one.

Instead, in the case of extra packet injection attack with no direct feedthrough, 
it can be shown that
 the likelihood function $\lambda(\Z_k|x_{k})$ can be written  as 
\begin{equation}\label{eq:likelihood:no}
\lambda(\Z_k|x_k) = \gamma(\Z_k) \, \left [ 1 - p_d + p_d \sum_{y_k \in \Z_k} \frac{\ell(y_{k}|x_{k})}{n \, \kappa(y_k)} \right]
\end{equation}
where $n$ denotes the cardinality of $\Z_k$, i.e. the number of received measurements.

Hence, the following result holds (the proof is omitted since it follows along the same lines as the proofs of Theorems 1 and 2).
\vspace{.3 cm}

\begin{theorem} (\textit{Correction without direct feedthrough})
	Suppose that the prior density 
	%	$p_{k|k-1}(\A_{k},x_{k}|\Z^{k-1})$ 
	at time $k$ is \textit{hybrid Bernoulli} of the form
	\be
		p(\A_{k-1},x_{k}|\Z^{k-1}) = \left\{ \ba{ll} 
		(1 - r_{k|k-1}) \, p^0_{k|k-1}(x_{k}),                   & \mbox{if } \A_{k-1} = \emptyset      \vspace{2mm} \\
		r_{k|k-1} \cdot p^1_{k|k-1}(a_{k-1},x_{k}),                   & \mbox{if } \A_{k-1} = \{a_{k-1}\}                                  
		\ea \right..
	\label{eq:prediction_0:no}
	\ee
	Then, given the measurement random set $\Z_k$ for packet substitution attack, 
	also the posterior density at time $k$ turns out to be \textit{hybrid Bernoulli} of the form
	\be
	p(\A_{k-1},x_k|\Z^k) = \left\{ \ba{ll} 
	(1 - r_{k|k}) \, p^0_{k|k}(x_k),                   & \mbox{if } \A_{k-1} = \emptyset      \vspace{2mm} \\
	r_{k|k} \cdot p^1_{k|k}(a_{k-1},x_k),                  & \mbox{if } \A_{k-1} = \{a_{k-1}\}                                  
	\ea \right.
	\label{eq:correction_final:no}
	\ee
	The triplet $\left ( r_{k|k} , p^0_{k|k}(x_k) , p^1_{k|k}(a_{k-1},x_k) \right )$ completely specifying the posterior density can be
	computed as in Theorem 1 for the case of packet substitution and as in Theorem 2  for the case of extra packet injection attack, provided that
	$a_k$, $\mathcal A_k$, and $\ell(y_k|a_k,x_k)$ are replaced by $a_{k-1}$, $\mathcal A_{k-1}$, and $\ell(y_k|x_k)$, respectively.
	\end{theorem}

\subsection{Dynamic model and prediction}\label{sec4.2}
The joint transition density takes the form
\be
\pi(\A_{k},x_{k+1}|\A_{k-1},x_k) = \pi(x_{k+1}|\A_k,x_k) \, \pi(\A_{k}|\A_{k-1}) 
\ee
where
\be
\pi(x_{k+1}|\A_k,x_k) = \left\{ \ba{ll} 
\pi(x_{k+1}|x_k),                  & \mbox{if } \A_k = \emptyset      \vspace{2mm} \\
\pi(x_{k+1}|a_k,x_k),                & \mbox{if } \A_k = \{a_k\}                                  
\ea \right.
\ee
with  $\pi(x_{k+1}|x_k)$ and $\pi(x_{k+1}|a_k,x_k)$ known Markov transition PDFs. 

The transition density $\pi(\A_{k}|\A_{k-1})$ of the attack BRS takes the form
\begin{eqnarray} 
	\pi(\A_{k}|\emptyset) &=& \left\{ \ba{ll} 
	1 - p_b,                   & \mbox{if } \A_{k} = \emptyset      \vspace{2mm} \\
	p_b \, p(a_{k}),                  & \mbox{if } \A_{k} = \{a_{k}\}                                  
	\ea \right. \nonumber
	\\	
	\pi(\A_{k}|\{a_{k-1}\}) &=& \left\{ \ba{ll} 
	1 - p_s,                  & \mbox{if } \A_{k} = \emptyset      \vspace{2mm} \\
	p_s \, p(a_{k}),                  & \mbox{if } \A_{k} = \{a_{k}\}                                  
	\ea \right. \nonumber
\end{eqnarray}
$p(a_{k})$ is the PDF summarizing the available knowledge on  $a_{k}$, which can be taken equal to an uninformative PDF 
(e.g., uniform over the attack space) when the attack vector is completely unknown.

%%%%%%%%%%%%%%%%%%%%%%%%%%%%%%%%%%%%%%%%%%%%%%%%%%%%%%%%%%%%%%%%%%%%%%%%%%%%%%%%%%%%%%%%%%%%%%%
%%%%%%%%%%%%%%%%%%%%%%%%%%%%%%%%%%%%%%%%%%%%%%%%%%%%%%%%%%%%%%%%%%%%%%%%%%%%%%%%%%%%%%%%%%%%%%%
\vspace{.3 cm}

\begin{theorem}
	Given the posterior hybrid Bernoulli density 
	$p(\A_{k-1},x_{k}|\Z^{k})$ 
	at time $k$ of the form (\ref{eq:correction_final:no}),
	fully characterized by the triplet $\big( r_{k|k}, p^0_{k|k}(x_k), p^1_{k|k}(a_{k-1},x_k) \big)$,
	also the predicted density turns out to be hybrid Bernoulli of the form 
	\begin{align} \label{eq:prediction_final:no}	
		&& p(\A_{k},x_{k+1}|\Z^{k}) =  \left\{ \ba{ll} 
		(1-r_{k+1|k}) \, p^0_{k+1|k}(x_{k+1}),                   &   \mbox{if } \A_{k} = \emptyset      \vspace{2mm}   \\
		r_{k+1|k} \cdot p^1_{k+1|k}(a_{k},x_{k+1}),                   &   \mbox{if } \A_{k} = \{a_{k}\}                                 
		\ea \right.  
	\end{align} 
	with
	\begin{align} 
		r_{k+1|k} = && (1 - r_{k|k}) \, p_b + r_{k|k} \, p_s  \label{p_existence1:no} 
\\
		p^0_{k+1|k}(x_{k+1}) =&&  \frac{(1 - r_{k|k}) (1 - p_b) \, p_{k+1|k}(x_{k+1}|\emptyset)}{1 - r_{k+1|k}} 
		+ \frac{r_{k|k} (1 - p_s) \, p_{k+1|k}(x_{k+1}|\{a_{k-1}\})}{1 - r_{k+1|k}}  \label{p_01:no} 
\\
		p^1_{k+1|k}(a_{k},x_{k+1}) = &&  \frac{(1 - r_{k|k}) \, p_b \, p_{k+1|k}(x_{k+1}|\{a_{k}\},\emptyset) \, p(a_{k})}{r_{k+1|k}}  + \frac{r_{k|k} \, p_s \, p_{k+1|k}(x_{k+1}|\{a_{k}\},\{a_{k-1}\}) \, p(a_{k})}{r_{k+1|k}} \label{p_11:no}
	\end{align} 
	where  
	\begin{align} 
		p_{k+1|k}(x_{k+1}|\emptyset) \defi&& \int \pi(x_{k+1}|x_{k}) \, p^0_{k|k}(x_{k}) \, \mbox{d} x_{k} 
\\
		p_{k+1|k}(x_{k+1}|\{a_{k-1}\}) \defi&& \iint \pi(x_{k+1}|x_{k}) \, p^1_{k|k}(a_{k-1},x_{k}) \,  \mbox{d} a_{k-1} \mbox{d} x_{k}   
\\
p_{k+1|k}(x_{k+1}|\{a_{k}\},\emptyset) \defi&& \int \pi(x_{k+1}|a_k,x_{k}) \, p^0_{k|k}(x_{k}) \, \mbox{d} x_{k}  
\\
		p_{k+1|k}(x_{k+1}|\{a_{k}\},\{a_{k-1}\}) \defi&& \iint \pi(x_{k+1}|a_k,x_{k}) \, p^1_{k|k}(a_{k-1},x_{k}) \,  \mbox{d} a_{k-1} \mbox{d} x_{k}    .
	\end{align} 	
	\normalsize
	%	\hfill $\square$
\end{theorem} 

\textit{ Proof:} The prediction equation is given by the following generalization of (\ref{eq:ax:no}) 
\begin{eqnarray} \label{eq:prediction_bayes}
p(\A_{k},x_{k+1}|\Z^{k}) && =  \iint \pi(\A_{k},x_{k+1}|\A_{k-1},x_{k}) 
\, p(\A_{k-1},x_{k}|\Z^{k}) \, \delta \A_{k-1} \mbox{d} x_{k}  \nonumber   \\
&&  = (1 - r_{k|k}) \int \pi(\A_{k},x_{k+1}|\emptyset,x_{k}) \, p^0_{k|k}(x_{k}) \, \mbox{d} x_{k}    \nonumber   \\
&&
 + \, r_{k|k} \iint \pi(\A_{k},x_{k+1}|\{a_{k-1}\},x_{k}) \, p^1_{k|k}(a_{k-1},x_{k}) \, \mbox{d} a_{k-1} \mbox{d} x_{k} \nonumber
\end{eqnarray} \normalsize  
Then, for $\A_{k} = \emptyset$, one has 
\begin{eqnarray*} 
p(\emptyset,x_{k+1}|\Z^{k}) &=& (1 - r_{k|k}) (1 - p_b) \int \pi(x_{k+1}|x_k) \, p^0_{k|k}(x_{k}) \, \mbox{d} x_{k}  
 \nonumber   \\
&& 
+ \, r_{k|k} (1 - p_s) \iint \pi(x_{k+1}|x_k) \, p^1_{k|k}(a_{k-1},x_{k}) \,  \mbox{d} a_{k-1} \mbox{d} x_{k} 
 \nonumber   \\
&=& (1 - r_{k|k}) \, (1 - p_b) \, p_{k+1|k}(x_{k+1}|\emptyset) 
+ \, r_{k|k} \, (1 - p_s) \, p_{k+1|k}(x_{k+1}|\{a_{k-1}\}) .
\label{eq:prediction_a01}
\end{eqnarray*}  
Analogously, for $\A_{k} = \{a_{k}\}$ we obtain
\begin{eqnarray*} 
p(\{a_{k}\},x_{k+1}|\Z^{k}) = \bigg[ (1 - r_{k|k}) \, p_b \, p_{k+1|k}(x_{k+1}|\{a_{k}\},\emptyset)
+ \, r_{k|k} \, p_s \, p_{k+1|k}(x_{k+1}|\{a_{k}\},\{a_{k-1}\}) \bigg] \, p(a_{k})  .
\label{eq:prediction_a1}
\end{eqnarray*}  
Thus, the output of the prediction step is of the form
(\ref{eq:prediction_final:no}), fully specified by (\ref{p_existence1:no})-(\ref{p_11:no}).

\section{Gaussian-mixture Hybrid Bernoulli filter}\label{sec5}

While in general no exact closed-form solution to the proposed hybrid Bernoulli filter is admitted, for the special class of linear Gaussian models,
this problem can be effectively mitigated by parameterizing the posterior densities $p^0_{k|k}(\cdot)$ and $p^1_{k|k}(\cdot,\cdot)$ via Gaussian mixtures (GMs)
so as to derive a GM hybrid Bernoulli filter.
This approach can be generalized to nonlinear models and/or non-Gaussian noises via nonlinear extensions of the GM approximation 
based on nonlinear filtering techniques such as the Extended Kalman Filter or the Unscented Kalman filter. 
In what follows, a detailed derivation of the GM hybrid Bernoulli filter for linear-Gaussian models is provided.
For the sake of brevity, only the direct feedthrough case (Section~\ref{sec3}) is considered. The GM implementation in the case of no direct feedthrough (Section~\ref{sec4}) can be derived in a similar way.

Denoting by $\mathcal{N}(x;m,P)$ a Gaussian PDF in the variable $x$, with mean $m$ and covariance $P$, the closed-form GM hybrid Bernoulli filter 
assumes linear Gaussian observation, transition, and (a priori) attack models, i.e.
\begin{eqnarray}  \label{eq:GM_like}
	\ell(y_k|x_k) &=& \mathcal{N} (y_k; C x_k, R)  \\ 
	\ell(y_k|a_k,x_k) &=& \mathcal{N} (y_k; C x_k + H a_k, R)  \label{eq:GM_like2}  \\ 
	\pi(x_{k+1}|x_k) &=& \mathcal{N}  (x_{k+1}; A x_k, Q)  \label{eq:GM_trans} \\  
	\pi(x_{k+1}|a_k,x_k) &=& \mathcal{N}  (x_{k+1}; A x_k + G a_k, Q) \label{eq:GM_trans2} \\
	p(a) &=& \sum_{j=1}^{J^a} \tilde{\omega}^{a,j} \mathcal{N} (a; \tilde{a}^j,\tilde{P}^{a,j}) 
	\label{eq:GM_attack}
\end{eqnarray} 
\color{black}
Note that \eqref{eq:GM_attack} uses given model parameters $J^a, \tilde{\omega}^{a,j}, \tilde{a}^j, \tilde{P}^{a,j}, j=1,\dots,J^a$, to define the a priori PDF of the signal attack, here expressed as a Gaussian mixture and supposed time independent.

In the GM implementation, each probability density at time $k$ is represented by the following set of parameters 
\begin{eqnarray} 
	\Big( r_{k|k}, p^{0}_{k|k}(x_k), p^{1}_{k|k}(a_k,x_k) \Big)   
	= \Big( r_{k|k}, \big\{ \omega^{0,j}_{k|k}, m^{0,j}_{k|k}, P^{0,j}_{k|k} \big\}^{J^0_{k|k}}_{j=1}, \big\{ \omega^{1,j}_{k|k}, m^{1,j}_{k|k}, P^{1,j}_{k|k} \big\}^{J^1_{k|k}}_{j=1} \Big)
\end{eqnarray} 
where
$\omega$ and $J$ indicate, respectively, weights and number of mixture components, such that
\begin{eqnarray} 
	p^{0}_{k|k}(x_k) 
	= \sum_{j=1}^{J^0_{k|k}} \omega^{0,j}_{k|k} \, \mathcal{N}(m^{0,j}_{k|k},P^{0,j}_{k|k}) 
	\label{p0kk}
	\\
	p^{1}_{k|k}(a_k,x_k) 
	= \sum_{j=1}^{J^1_{k|k}} \omega^{1,j}_{k|k} \, \mathcal{N}(m^{1,j}_{k|k},P^{1,j}_{k|k}) 
	\label{p1kk}
\end{eqnarray} 
%where we defined
%$m^0_{k|k} = \hat{x}_{k|k}$ and $m^{1,ij}_{k|k} = \begin{bmatrix} 
%\hat{x}^{1,ij}_{k|k} \\
%\hat{a}^{ij}_k
%\end{bmatrix}  $
with
$m^0_{k|k} = \hat{x}^0_{k|k}$, 
$m^{1}_{k|k} = [ \hat{x}^{1^T}_{k|k} , \hat{a}_k^T ]^T$, $P^{0}_{k|k} \defi \mathbb{E}[(x_k - \hat{x}^{0}_{k|k}) (x_k - \hat{x}^{0}_{k|k})^T]$, $P^{1}_{k|k} = \begin{bmatrix} 
P^{1x}_{k|k} & P^{xa}_k \\
P^{ax}_k & P^{a}_k \end{bmatrix}$,
and $P^{1x}_{k|k} \defi \mathbb{E}[(x_k - \hat{x}^{1}_{k|k}) (x_k - \hat{x}^{1}_{k|k})^T]$,
$(P^{xa}_k)^T = P^{ax}_k \defi \mathbb{E}[(a_k - \hat{a}_{k}) (x_k - \hat{x}^{1}_{k|k})^T]$,
$P^{a}_k \defi \mathbb{E}[(a_k - \hat{a}_{k}) (a_k - \hat{a}_{k})^T]$.
The weights are such that 
$\sum_{j=1}^{J^{0}_{k|k}} \omega^{0,j}_{k|k} = 1$, and $\sum_{j=1}^{J^{1}_{k|k}} \omega^{1,j}_{k|k} = 1$.

The Gaussian Mixture implementation of the \textit{Hybrid Bernoulli Filter} (GM-HBF) is described as follows.

%%%%%%%%%%%%%%%%%%%%%%%%%%%%%%%%%%%
%\vspace{.3 cm}
\subsection{GM-HBF correction for packet substitution}\label{sec5.1}

\begin{proposition} 
	Suppose that: 
	assumptions \eqref{eq:GM_like}-\eqref{eq:GM_attack} hold; 
	the measurement set $\Z_k$ is defined by \eqref{ps-attack};
	the predicted FISST density at time $k$ is fully specified by the triplet $\big( r_{k|k-1}, p^0_{k|k-1}(x_k), p^1_{k|k-1}(a_k,x_k) \big)$;
	$p^0_{k|k-1}(\cdot)$, $p^1_{k|k-1}(\cdot,\cdot)$ are Gaussian mixtures of the form 
	\begin{eqnarray} 
		p^{0}_{k|k-1}(x_{k}) && \hspace{-.6cm} = \sum_{j=1}^{J^{0}_{k|k-1}} \omega^{0,j}_{k|k-1} \mathcal{N} (m^{0,j}_{k|k-1}, P^{0,j}_{k|k-1}) \label{inizp0}
		\\
		p^{1}_{k|k-1}(a_k,x_k) &=& \sum_{j=1}^{J^{1}_{k|k-1}} \omega^{1,j}_{k|k-1} \mathcal{N} (m^{1,j}_{k|k-1}, P^{1,j}_{k|k-1})
		\label{inizp1} 
	\end{eqnarray} 
	%	where $m^{0,j}_{k|k-1} = \hat{x}^{0,j}_{k|k-1}$, $m^{1,j}_{k|k-1} = \hat{x}^{1,j}_{k|k-1}$, $\sum_{j=1}^{J^{0}_{k|k-1}} \omega^{0,j}_{k|k-1} = 1$, and $\sum_{j=1}^{J^{1}_{k|k-1}} \omega^{1,j}_{k|k-1} = 1$.
	Then, the posterior FISST density $\big( r_{k|k}, p^0_{k|k}(x_k), p^1_{k|k}(a_k,x_k) \big)$ 
	%	under the \textit{packet substitution} measurement model 
	is given by  
	\small
	\begin{eqnarray} 
		%			&&\hspace{-2.5cm}
		r_{k|k} &=& \frac{(1 - p_f) \, \Psi_1 + p_f \kappa(y_k)}{(1 - p_f) (\Psi_0 - r_{k|k-1} \Psi) + p_f \kappa(y_k)} \, r_{k|k-1}  \label{rkkprop1}
		%		r_{k|k} &=& \frac{1 - p_d + p_d \, \Gamma_1}{1 - p_d + p_d(1 - r_{k|k-1}) \Gamma_0 + p_d \, r_{k|k-1} \Gamma_1} \, r_{k|k-1}   
		\\
		p^{0}_{k|k}(x_k) 
		&=& 
		\sum_{j=1}^{J^{0}_{k|k}} \omega^{0,j}_{k|k} \mathcal{N} (m^{0,j}_{k|k}, P^{0,j}_{k|k}) =
		\sum_{j=1}^{J^{0}_{k|k-1}} \omega^{0,j}_{F,k|k} \, \mathcal{N} (m^{0,j}_{k|k-1}, P^{0,j}_{k|k-1}) 
		+ \sum_{j=1}^{J^{0}_{k|k-1}} \omega^{0,j}_{\bar{F},k|k} \, \mathcal{N} (m^{0,j}_{k|k}, P^{0,j}_{k|k}) \nonumber
		\\
		\label{p0kps}
		\\
		p^{1}_{k|k}(a_k,x_k)
		&=&
		\sum_{j=1}^{J^{1}_{k|k}} \omega^{1,j}_{k|k} \mathcal{N} (m^{1,j}_{k|k}, P^{1,j}_{k|k}) =
		\sum_{j=1}^{J^{1}_{k|k-1}} \omega^{1,j}_{F,k|k} \, \mathcal{N} (m^{1,j}_{k|k-1}, P^{1,j}_{k|k-1})   
		+ \sum_{j=1}^{J^{1}_{k|k-1}} \omega^{1,j}_{\bar{F},k|k} \, \mathcal{N} (m^{1,j}_{k|k}, P^{1,j}_{k|k}) \nonumber
\\		\label{p1kps}
	\end{eqnarray} \normalsize
	where 
	\begin{eqnarray} 
		\omega^{i,j}_{F,k|k} &=& \frac{p_f \, \kappa(y_k) \, \omega^{i,j}_{k|k-1}}{(1-p_f) \Psi_i + p_f \kappa(y_k)}, \label{wf00} \\
		\omega^{i,j}_{\bar{F},k|k} &=& \frac{(1-p_f) \, \omega^{i,j}_{k|k-1}}{(1-p_f) \Psi_i + p_f \, \kappa(y_k)} \, q^{i,j}_k (y_k) \label{wf01}
		%		\\
		%		\omega^{1,j}_{F,k|k} = \frac{p_f \, \kappa(y_k) \, \omega^{1,j}_{k|k-1}}{(1-p_f) \Psi_1 + p_f \kappa(y_k)}, \quad  
		%		\omega^{1,j}_{\bar{F},k|k} = \frac{(1-p_f) \, \omega^{1,j}_{k|k-1}}{(1-p_f) \Psi_1 + p_f \, \kappa(y_k)} \, q^{1,j}_k (y_k) \label{wf1}
	\end{eqnarray} 
	for $i=0,1$,
	while
	\begin{eqnarray} 
		q^{0,j}_k (y_k) &=& \mathcal{N} (y; C m^{0,j}_{k|k-1}, C P^{0,j}_{k|k-1} C^T + R )
		\label{q0j}
		\\
		q^{1,j}_k (y_k) &=& \mathcal{N} (y; \tilde{C} m^{1,j}_{k|k-1}, \tilde{C} P^{1,j}_{k|k-1} \tilde{C}^T + R ) \label{q1j}
	\end{eqnarray} 
	with $\tilde{C} \defi [C, H]$, $\Psi_0 = \sum_{j=1}^{J^{0}_{k|k-1}} \omega^{0,j}_{k|k-1} q^{0,j}_k (y_k)$, and $\Psi_1 = \sum_{j=1}^{J^{1}_{k|k-1}} \omega^{1,j}_{k|k-1} q^{1,j}_k (y_k)$.
	
\end{proposition}
%\vspace{.3 cm}
%{\em Proof:}

\textit{Proof:}
From Theorem 1, the corrected probability of signal attack existence is provided by \eqref{eq:correction_r} 
%as 
%\be
%r_{k|k} = \frac{(1 - p_f) \, \Psi_1 + p_f \kappa(y_k)}{(1 - p_f) (\Psi_0 - r_{k|k-1} \Psi) + p_f \kappa(y_k)} \, r_{k|k-1}
%\label{rkk_ps}
%\ee
where $\Psi_0$ is obtained by substituting \eqref{eq:GM_like} and \eqref{inizp0} into \eqref{psi_0}, so that 
\be
\Psi_0 = 
\displaystyle 
\int \mathcal{N} (y; C x_k, R)  \sum_{j=1}^{J^{0}_{k|k-1}} \omega^{0,j}_{k|k-1} \mathcal{N} (m^{0,j}_{k|k-1}, P^{0,j}_{k|k-1}) \, \mbox{d} x_{k}  .
\label{gmpsi0}
\ee
Then, by applying a standard result for Gaussian functions, \cite[Lemma 1]{Vo2006} we can write
\begin{eqnarray} 
	\int \mathcal{N} (y; C x_k, R) \, \mathcal{N} (m^{0,j}_{k|k-1}, P^{0,j}_{k|k-1}) \, \mbox{d} x_{k}
	= q^{0,j}_k (y_k) 
	%\, \mathcal{N} (m^{0,j}_{k|k}, P^{0,j}_{k|k})   
	\label{gmqn}
\end{eqnarray}
where $q^{0,j}_k (y_k)$ is given by (\ref{q0j})
% \be
%q^{0,j}_k (y_k) = \mathcal{N} (y; C m^{0,j}_{k|k-1}, C P^{0,j}_{k|k-1} C^T + R )
%\label{q0_def}
%\ee
and, hence, \eqref{gmpsi0} takes the form
\be
\Psi_0 = 
\sum_{j=1}^{J^{0}_{k|k-1}} \omega^{0,j}_{k|k-1} q^{0,j}_k (y_k) .
\ee
Moreover, $\Psi_1$ in \eqref{rkkprop1} can be analogously obtained by substituting \eqref{eq:GM_like2} and \eqref{inizp1} into \eqref{psi_1}, 
and by applying Lemma 1 in Vo and Ma \cite{Vo2006} to the (double) integral 
$\iint \mathcal{N} (y; C x_k + H a_k, R) \, \mathcal{N} (m^{1,j}_{k|k-1}, P^{1,j}_{k|k-1}) \, \mbox{d} a_{k} \mbox{d} x_{k}$, so as to obtain 
%i.e.
\be
\Psi_1 = \sum_{j=1}^{J^{1}_{k|k-1}} \omega^{1,j}_{k|k-1} q^{1,j}_k (y_k)
\label{gmgamma1}
\ee
where $q^{1,j}(y_k)$ is given by (\ref{q1j}) 
%\be
%q^{1,j}_k (y_k) = \mathcal{N} (y; \tilde{C} m^{1,j}_{k|k-1}, \tilde{C} P^{1,j}_{k|k-1} \tilde{C}^T + R )  
%\label{q1_def}
%\ee
and $m^{1,j}_{k|k-1} = [ (\hat{x}^{1}_{k|k-1})^T, (\hat{a}^j_k)^T ]^T$.

Next, the posterior density $p^{0}_{k|k}(\cdot)$ can be derived from \eqref{eq:correction_0}
in Theorem 1 as
\begin{eqnarray} 
	p^{0}_{k|k}(x_k) = \frac{p_f \, \kappa(y_k)}{(1 - p_f) \, \Psi_0 + p_f \, \kappa(y_k)} \, p^{0}_{k|k-1}(x_k)  
	+ \frac{(1 - p_f) \, \ell(y_k|x_k)}{(1 - p_f) \, \Psi_0 + p_f \, \kappa(y_k)} \, p^{0}_{k|k-1}(x_k)  .
	\label{theo1ps} 
\end{eqnarray}
By substituting \eqref{eq:GM_like} and \eqref{inizp0} into \eqref{theo1ps}, we obtain 
\begin{eqnarray} 
	p^{0}_{k|k}(x_k) &=& \sum_{j=1}^{J^{0}_{k|k-1}} \frac{p_f \, \kappa(y_k) \, \omega^{0,j}_{k|k-1}}{(1 - p_f) \, \Psi_0 + p_f \, \kappa(y_k)} \, \mathcal{N} (m^{0,j}_{k|k-1}, P^{0,j}_{k|k-1})      
	\nonumber    \\
	 && + \sum_{j=1}^{J^{0}_{k|k-1}} \frac{(1 - p_f) \, \omega^{0,j}_{k|k-1} \, \mathcal{N} (y; C x_k, R)}{(1 - p_f) \, \Psi_0 + p_f \, \kappa(y_k)} 
	\, \mathcal{N} (m^{0,j}_{k|k-1}, P^{0,j}_{k|k-1}) 	  . 
	\label{eq:gmp0cps}
\end{eqnarray}
Then, by applying Lemma 2 in Vo and Ma, \cite{Vo2006} we can write
\begin{eqnarray} 
	\mathcal{N} (y; C x_k, R) \, \mathcal{N} (m^{0,j}_{k|k-1}, P^{0,j}_{k|k-1}) 
	= q^{0,j}_k (y_k) \, \mathcal{N} (m^{0,j}_{k|k}, P^{0,j}_{k|k})  
	\label{gmqn_ps}
\end{eqnarray}
where $q^{0,j}_k (y_k)$ has been defined in \eqref{q0j}, while $m^{0,j}_{k|k}, P^{0,j}_{k|k}$ have been introduced in \eqref{p0kk}.

In the special case of linear Gaussian models, 
%the correction steps can be carried out as in the JISE filter 
$m^{0,j}_{k|k}$ and $P^{0,j}_{k|k}$ can be easily calculated following the standard Bayes filter correction step, which in this case boils down to the standard Kalman filter for linear discrete-time systems \cite{Gillijns2007}:  
%with direct feedthrough. Here we followed the steps in
\begin{eqnarray} 
	m^{0,j}_{k|k} &=& m^{0,j}_{k|k-1} + L^{0,j}_k (y_k - C m^{0,j}_{k|k-1})   \label{cgmx0_ps}   \\
	P^{0,j}_{k|k}  &=& (I - L^{0,j}_k C) P^{0,j}_{k|k-1} \label{cgmP0_ps}
\end{eqnarray}
where
\begin{eqnarray} 
	L^{0,j}_k &=& P^{0,j}_{k|k-1} {C}^{T} (S^{0,j}_{k})^{-1}       \\
	S^{0,j}_{k} &=& C P^{0,j}_{k|k-1} {C}^{T} + R   . \label{S0}
\end{eqnarray}
%Thus, we can rewrite \eqref{eq:gmp1c} as
Thus, by substituting \eqref{gmqn_ps} into \eqref{eq:gmp0cps} with means and covariances given by \eqref{cgmx0_ps}-\eqref{cgmP0_ps}, we can write 
\be
p^{0}_{k|k}(x_k) = \sum_{j=1}^{J^{0}_{k|k}} \omega^{0,j}_{k|k} \, \mathcal{N} (m^{0,j}_{k|k}, P^{0,j}_{k|k})
\ee
which consists of $2 \, J^{0}_{k|k-1}$ Gaussian components, i.e.
\begin{eqnarray} 
	p^{0}_{k|k}(x_k) = \sum_{j=1}^{J^{0}_{k|k-1}} \omega^{0,j}_{F,k|k} \, \mathcal{N} (m^{0,j}_{k|k-1}, P^{0,j}_{k|k-1}) 
	+ \sum_{j=1}^{J^{0}_{k|k-1}} \omega^{0,j}_{\bar{F},k|k} \, \mathcal{N} (m^{0,j}_{k|k}, P^{0,j}_{k|k})
	\label{1+z_ps} 
\end{eqnarray}
with weights $\omega_{F,k|k}^{0,j}, \omega_{\bar{F},k|k}^{0,j}$ given by (\ref{wf00})-(\ref{wf01}) for $i=0$.
%given by \eqref{wab1}.
%\begin{eqnarray} 
%	\omega^{0,j}_{F,k|k} &=& \frac{p_f \, \kappa(y_k) \, \omega^{0,j}_{k|k-1}}{(1 - p_f) \, \Psi_0 + p_f \, \kappa(y_k)}   \\
%	\omega^{0,j}_{\bar{F},k|k} &=& \frac{(1 - p_f) \, \omega^{0,j}_{k|k-1} q^{0,j}_k (y_k)}{(1 - p_f) \, \Psi_0 + p_f \, \kappa(y_k)}    .
%\end{eqnarray}
Note that, as it can be seen from \eqref{1+z_ps}, it turns out that $J^{0}_{k|k} = 2 \,  J^{0}_{k|k-1}$, where the first \textit{legacy} (not corrected) components correspond to the hypothesis of the system-originated measurement being replaced by a fake one $y_k^f$, while the remaining components are the ones corrected under the hypothesis of receiving $y_k$ with probability $1 - p_f$.

Following the same rationale, analogous results can be obtained for $p^{1}_{k|k}(\cdot,\cdot)$, with the exception that also signal attack estimation has to be performed. 
By substituting \eqref{eq:GM_like2} and \eqref{inizp1} into \eqref{eq:correction_1} in Theorem 1, we obtain 
\begin{eqnarray} 
p^{1}_{k|k}(a_k,x_k)
&=& \sum_{j=1}^{J^{1}_{k|k-1}} \frac{p_f \, \kappa(y_k) \, \omega^{1,j}_{k|k-1}}{(1 - p_f) \, \Psi_1 + p_f \, \kappa(y_k)} \mathcal{N} (m^{1,j}_{k|k-1}, P^{1,j}_{k|k-1}) 
\nonumber
\\
&& + \sum_{j=1}^{J^{1}_{k|k-1}} \frac{(1 - p_f) \, \omega^{1,j}_{k|k-1} \, \mathcal{N} (y; C x_k + H a_k, R)}{(1 - p_f) \, \Psi_1 + p_f \, \kappa(y_k)} 
\mathcal{N} (m^{1,j}_{k|k-1}, P^{1,j}_{k|k-1}) 	  . 
\label{eq:gmp1cps}
\end{eqnarray}
\normalsize
Then, by applying Lemma 2 in Vo and Ma, \cite{Vo2006} we can write
\begin{eqnarray} 
	\mathcal{N} (y; C x_k + H a_k, R) \, \mathcal{N} (m^{1,j}_{k|k-1}, P^{1,j}_{k|k-1}) 
	= q^{1,j}_k (y_k) \, \mathcal{N} (m^{1,j}_{k|k}, P^{1,j}_{k|k})  
	\label{gmqn1_ps}
\end{eqnarray}
where $q^{1,j}_k (y_k)$ has been defined in \eqref{q1j}, while $m^{1,j}_{k|k}, P^{1,j}_{k|k}$ have been introduced in \eqref{p1kk}.
For linear Gaussian models, 
%the correction steps can be carried out as in the JISE filter 
$m^{1,j}_{k|k}$ and $P^{1,j}_{k|k}$ can be calculated following the correction step of the filter for joint input and state estimation
of linear discrete-time systems \cite{Gillijns2007}, 
%with direct feedthrough. Here we followed the steps in
 introduced in Section~\ref{sec2.2}. In particular, $m^{1,j}_{k|k}$ consists of:
\begin{eqnarray} 	
	\hat{x}^{1,j}_{k|k} &=& \hat{x}^{1,j}_{k|k-1} + \tilde{L}^{1,j}_k (y_k - C \hat{x}^{1,j}_{k|k-1} - H \hat{a}^{j}_k) 
	= \hat{x}^{1,j}_{k|k-1} + L^{1,j}_k (y_k - C \hat{x}^{1,j}_{k|k-1}) \label{cgmx1_ps}   \\
	\hat{a}^{j}_k &=& M^{j}_k (y_k - C \hat{x}^{1,j}_{k|k-1})   \label{cgma_ps} 
\end{eqnarray}
where
\begin{eqnarray} 
	L^{1,j}_k &=& \tilde{L}^{1,j}_k (I - H M^{j}_k)       \\
	\tilde{L}^{1,j}_k &=& P^{1x,j}_{k|k-1} {C}^{T} (S^{1,j}_{k})^{-1}       \\
	S^{1,j}_{k} &=& C P^{1x,j}_{k|k-1} {C}^{T} + R \\
	M^{j}_k &=& \big[ H^T (S^{1,j}_{k})^{-1} H \big]^{-1} {H}^{T} (S^{1,j}_{k})^{-1}    .
\end{eqnarray}
The elements composing $P^{1,j}_{k|k}$ can be computed as 
\begin{eqnarray} 
	P^{1x,j}_{k|k}  &=& (I - L^{1,j}_k C) P^{1x,j}_{k|k-1}  \label{cP_ps} \\
	P^{a,j}_k &=& [ {H}^{T} (S^{1,j}_{k})^{-1} H ]^{-1}    \label{cPa_ps} \\
	%P^{1x,j}_{k|k}  &=& P^{1,j}_{k|k-1} - L^{1,j}_k ( \tilde{R}^{j}_{k} - H P^{a,j}_k {H}^{T} )  {L^{1,j}_k}^{T}  \\
	P^{xa,j}_k &=&  (P^{ax,j}_k)^{T} = - \tilde{L}^{1,j}_k H P^{a,j}_k     \label{cPxa_ps}   .
\end{eqnarray}
%Thus, we can rewrite \eqref{eq:gmp1c} as
Thus, by substituting \eqref{gmqn1_ps} into \eqref{eq:gmp1cps} with means and covariances given by \eqref{cgmx1_ps}-\eqref{cgma_ps} and \eqref{cP_ps}-\eqref{cPxa_ps}, we can write 
\be
p^{1}_{k|k}(a_k,x_k) = \sum_{j=1}^{J^{1}_{k|k}} \omega^{1,j}_{k|k} \mathcal{N} (m^{1,j}_{k|k}, P^{1,j}_{k|k})   
\ee
which comprises $2 \, J^{1}_{k|k-1}$ components, i.e.
\begin{eqnarray} 
	p^{1}_{k|k}(a_k,x_k) = \sum_{j=1}^{J^{1}_{k|k-1}} \omega^{1,j}_{F,k|k} \, \mathcal{N} (m^{1,j}_{k|k-1}, P^{1,j}_{k|k-1}) 
	+ \sum_{j=1}^{J^{1}_{k|k-1}} \omega^{1,j}_{\bar{F},k|k} \, \mathcal{N} (m^{1,j}_{k|k}, P^{1,j}_{k|k})
	\label{1+z_1ps} 
\end{eqnarray}
with weights $\omega_{F,k|k}^{1,j}, \omega_{\bar{F},k|k}^{1,j}$ given by (\ref{wf00})-(\ref{wf01}) for $i=1$.%given by \eqref{wab1}.
%\begin{eqnarray} 
%	\omega^{1,j}_{F,k|k} &=& \frac{p_f \, \kappa(y_k) \, \omega^{1,j}_{k|k-1}}{(1 - p_f) \, \Psi_1 + p_f \, \kappa(y_k)}   \\
%	\omega^{1,j}_{\bar{F},k|k} &=& \frac{(1 - p_f) \, \omega^{1,j}_{k|k-1} q^{1,j}_k (y_k)}{(1 - p_f) \, \Psi_1 + p_f \, \kappa(y_k)}    .  
%\end{eqnarray} 

% % % % % % % % % % % % % % % % % %
%\vspace{.3 cm}
\subsection{GM-HBF correction for extra packet injection}\label{sec5.2}

\begin{proposition} 
	Suppose that: 
	assumptions \eqref{eq:GM_like}-\eqref{eq:GM_attack} hold; 
	the measurement set $\Z_k$ is defined by \eqref{epi-attack};
	the predicted FISST density at time $k$ is fully specified by the triplet 
	$\big( r_{k|k-1}, p^0_{k|k-1}(x_k), p^1_{k|k-1}(a_k,x_k) \big)$;
	$p^0_{k|k-1}(\cdot)$, $p^1_{k|k-1}(\cdot,\cdot)$ are Gaussian mixtures of the form \eqref{inizp0} and \eqref{inizp1}, respectively.
	%	where $m^{0,j}_{k|k-1} = \hat{x}^{0,j}_{k|k-1}$, $m^{1,j}_{k|k-1} = \hat{x}^{1,j}_{k|k-1}$, $\sum_{j=1}^{J^{0}_{k|k-1}} \omega^{0,j}_{k|k-1} = 1$, and $\sum_{j=1}^{J^{1}_{k|k-1}} \omega^{1,j}_{k|k-1} = 1$.
	Then, the posterior FISST density $\big( r_{k|k}, p^0_{k|k}(x_k), p^1_{k|k}(a_k,x_k) \big)$ is given by  
	\small
	\begin{eqnarray} 
		r_{k|k} &=& \frac{1 - p_d + p_d \, \Gamma_1}{1 - p_d + p_d(1 - r_{k|k-1}) \, \Gamma_0 + p_d \, r_{k|k-1} \Gamma_1} \, r_{k|k-1}  
		\\
		p^{0}_{k|k}(x_k) 
		&=& 
		\sum_{j=1}^{J^{0}_{k|k}} \omega^{0,j}_{k|k} \mathcal{N} (m^{0,j}_{k|k}, P^{0,j}_{k|k}) 
		=
		\sum_{j=1}^{J^{0}_{k|k-1}} \omega^{0,j}_{\bar{D},k|k} \, \mathcal{N} (m^{0,j}_{k|k-1}, P^{0,j}_{k|k-1}) 
		+ \displaystyle \sum_{y_k \in \Z_k} \sum_{j=1}^{J^{0}_{k|k-1}} \omega^{0,j}_{D,k|k} \, \mathcal{N} (m^{0,j}_{k|k}, P^{0,j}_{k|k})  
		\nonumber \\
		\label{11}
		\\
		p^{1}_{k|k}(a_k,x_k)
		&=&
		\sum_{j=1}^{J^{1}_{k|k}} \omega^{1,j}_{k|k} \mathcal{N} (m^{1,j}_{k|k}, P^{1,j}_{k|k}) 
		=
		\sum_{j=1}^{J^{1}_{k|k-1}} \omega^{1,j}_{\bar{D},k|k} \, \mathcal{N} (m^{1,j}_{k|k-1}, P^{1,j}_{k|k-1})   
		+ \displaystyle \sum_{y_k \in \Z_k} \sum_{j=1}^{J^{1}_{k|k-1}} \omega^{1,j}_{D,k|k} \, \mathcal{N} (m^{1,j}_{k|k}, P^{1,j}_{k|k})  
				\nonumber \\
		\label{22}
	\end{eqnarray} 
	where, for $i=0,1$, 
	\begin{eqnarray} 
		\omega^{i,j}_{\bar{D},k|k} &=& \frac{(1-p_d) \, \omega^{i,j}_{k|k-1}}{1 - p_d + p_d \, \Gamma_i}, \\ 
		\omega^{i,j}_{D,k|k} &=& \frac{p_d \, \omega^{i,j}_{k|k-1} q^{i,j}_k (y_k)}{(1 - p_d + p_d \, \Gamma_i) \, n \, \kappa(y_k)} \label{wab01}
	\end{eqnarray} 
	and 
	\begin{eqnarray} 
		\Gamma_0 &=& \sum_{y_k \in \Z_k} \sum_{j=1}^{J^{0}_{k|k-1}} \frac{\omega^{0,j}_{k|k-1}}{n \, \kappa(y_k)} q^{0,j}_k (y_k) \label{gamma0_epi}
		\\
		\Gamma_1 &=& \sum_{y_k \in \Z_k} \sum_{j=1}^{J^{1}_{k|k-1}} \frac{\omega^{1,j}_{k|k-1}}{n \, \kappa(y_k)} q^{1,j}_k (y_k)  \label{gamma1_epi}.
	\end{eqnarray} 
\end{proposition}
\normalsize
%\vspace{.3 cm}
%{\em Proof:}

\textit{Proof:}
We first derive the corrected probability of signal attack existence, which can be directly written from \eqref{p_existence3} as 
\be
r_{k|k} = \frac{1 - p_d + p_d \, \Gamma_1}{1 - p_d + p_d(1 - r_{k|k-1}) \, \Gamma_0 + p_d \, r_{k|k-1} \Gamma_1} \, r_{k|k-1}   
\label{rkk}
\ee
where $\Gamma_0$ is obtained by substituting \eqref{eq:GM_like} and \eqref{inizp0} into \eqref{gamma_0}, so that 
\be
\Gamma_0 = 
\displaystyle 
\sum_{y_k \in \Z_k} \frac{\displaystyle \int \mathcal{N} (y; C x_k, R)  \sum_{j=1}^{J^{0}_{k|k-1}} \omega^{0,j}_{k|k-1} \mathcal{N} (m^{0,j}_{k|k-1}, P^{0,j}_{k|k-1}) \, \mbox{d} x_{k}}{n \, \kappa(y_k)}   .
\label{gmgamma0}
\ee
Then, by applying \eqref{gmqn}, 
\eqref{gmgamma0} takes the form \eqref{gamma0_epi}.
Moreover, $\Gamma_1$ in \eqref{rkk} can be analogously obtained by substituting \eqref{eq:GM_like2} and \eqref{inizp1} into \eqref{gamma_1}, 
and by applying \eqref{gmqn1_ps}
which leads to \eqref{gamma1_epi}.

Next, the posterior density $p^{0}_{k|k}(\cdot)$ can be derived from \eqref{p_03} in
Theorem 2 as
\begin{eqnarray} 
	p^{0}_{k|k}(x_k) = \frac{1 - p_d}{1 - p_d + p_d \Gamma_0} \, p^{0}_{k|k-1}(x_k)   
	+ \frac{p_d}{1 - p_d + p_d \Gamma_0} \, \displaystyle \sum_{y_k \in \Z_k} \frac{\ell(y_k|x_k)}{n \, \kappa(y_k)} p^{0}_{k|k-1}(x_k)  .
	\label{theo2} 
\end{eqnarray}
By substituting \eqref{eq:GM_like} and \eqref{inizp0} into (\ref{theo2}), we obtain 
\begin{eqnarray} 
	p^{0}_{k|k}(x_k) &=& \sum_{j=1}^{J^{0}_{k|k-1}} \frac{1 - p_d}{1 - p_d + p_d \Gamma_0} \, \omega^{0,j}_{k|k-1} \mathcal{N} (m^{0,j}_{k|k-1}, P^{0,j}_{k|k-1})        
			\nonumber \\
&& + \displaystyle \sum_{y_k \in \Z_k} \sum_{j=1}^{J^{0}_{k|k-1}} \omega^{0,j}_{k|k-1} \frac{p_d}{1 - p_d + p_d \Gamma_0} 
	\frac{\mathcal{N} (y; C x_k, R)}{n \, \kappa(y_k)} 
	\mathcal{N} (m^{0,j}_{k|k-1}, P^{0,j}_{k|k-1}) 	  .
	\label{eq:gmp0c}
\end{eqnarray}
Thus, by substituting \eqref{gmqn} into \eqref{eq:gmp0c}, with means and covariances given by \eqref{cgmx0_ps}-\eqref{cgmP0_ps}, we can write 
\be
p^{0}_{k|k}(x_k) = \sum_{j=1}^{J^{0}_{k|k}} \omega^{0,j}_{k|k} \mathcal{N} (m^{0,j}_{k|k}, P^{0,j}_{k|k})
\ee
which comprises $J^{0}_{k|k-1} (1 + |\Z_k|)$ components, where $|\Z_k|$ denotes the cardinality of the measurement set $\Z$ at time $k$, i.e.
\begin{eqnarray} 
	p^{0}_{k|k}(x_k) = \sum_{j=1}^{J^{0}_{k|k-1}} \omega^{0,j}_{\bar{D},k|k} \, \mathcal{N} (m^{0,j}_{k|k-1}, P^{0,j}_{k|k-1}) 
	+ \displaystyle \sum_{y_k \in \Z_k} \sum_{j=1}^{J^{0}_{k|k-1}} \omega^{0,j}_{D,k|k} \, \mathcal{N} (m^{0,j}_{k|k}, P^{0,j}_{k|k})
	\label{1+z} 
\end{eqnarray}
with weights 
%given by \eqref{wab1}.
\begin{eqnarray}  
	\omega^{0,j}_{\bar{D},k|k} &=& \frac{(1-p_d) \, \omega^{0,j}_{k|k-1}}{1-p_d+p_d \displaystyle \sum_{y_k \in \Z_k} \sum_{h=1}^{J^{0}_{k|k-1}} \frac{\omega^{0,h}_{k|k-1}}{n \, \kappa(y_k)} q^{0,h}_k (y_k)} \nonumber  \\
	\omega^{0,j}_{D,k|k} &=& \frac{p_d \, \omega^{0,j}_{k|k-1} q^{0,j}_k (y_k)}{\Big[  1-p_d+p_d \displaystyle \sum_{y_k \in \Z_k} \sum_{h=1}^{J^{0}_{k|k-1}} \frac{\omega^{0,h}_{k|k-1}}{n \, \kappa(y_k)} q^{0,h}_k (y_k) \Big]  \, n \, \kappa(y_k)}   \nonumber  .
\end{eqnarray}
Note that, as it can be seen from \eqref{1+z}, it turns out that $J^{0}_{k|k} = J^{0}_{k|k-1} + |\Z_k| \, J^{0}_{k|k-1}  = J^{0}_{k|k-1} (1 + |\Z_k|)$, where the first \textit{legacy} components correspond to the fact that no measurement has been delivered and hence no update is carried out, while the remaining components are the ones corrected when one or multiple measurements are received.

\noindent Following the same rationale, analogous results can be obtained for $p^{1}_{k|k}(\cdot,\cdot)$. 
From \eqref{p_13} in Theorem 2:
\begin{eqnarray} 
	p^{1}_{k|k}(a_k,x_k) = \frac{1 - p_d}{1 - p_d + p_d \Gamma_1} \, p^{1}_{k|k-1}(a_k,x_k)   
	+ \frac{p_d}{1 - p_d + p_d \Gamma_1} \, \displaystyle \sum_{y_k \in \Z_k} \frac{\ell(y_k|a_k,x_k)}{n \, \kappa(y_k)} p^{1}_{k|k-1}(a_k,x_k)  .
	\label{theo21}
\end{eqnarray}
By substituting \eqref{eq:GM_like2} and \eqref{inizp1} into (\ref{theo21}), we obtain 
\begin{eqnarray} 
	p^{1}_{k|k}(a_k,x_k) &=& \sum_{j=1}^{J^{1}_{k|k-1}} \frac{1 - p_d}{1 - p_d + p_d \Gamma_1} \, \omega^{1,j}_{k|k-1} \mathcal{N} (m^{1,j}_{k|k-1}, P^{1,j}_{k|k-1}) \label{eq:gmp1c}  \\ 
	&& + \displaystyle \sum_{y_k \in \Z_k} \sum_{j=1}^{J^{1}_{k|k-1}} \omega^{1,j}_{k|k-1} \frac{p_d}{1 - p_d + p_d \Gamma_1} 	
	\, \frac{\mathcal{N} (y; C x_k + H a_k, R)}{n \, \kappa(y_k)} \, 
	\mathcal{N} (m^{1,j}_{k|k-1}, P^{1,j}_{k|k-1}) 	. 
	\nonumber
\end{eqnarray}
Thus, by substituting \eqref{gmqn1_ps} into \eqref{eq:gmp1c}, with means and covariances given by \eqref{cgmx1_ps}-\eqref{cgma_ps} and \eqref{cP_ps}-\eqref{cPxa_ps}, we can write 
\be
p^{1}_{k|k}(a_k,x_k) = \sum_{j=1}^{J^{1}_{k|k}} \omega^{1,j}_{k|k} \mathcal{N} (m^{1,j}_{k|k}, P^{1,j}_{k|k})   
\ee
which comprises $J^{1}_{k|k-1} (1 + |\Z_k|)$ components, i.e.
\begin{eqnarray} 
	p^{1}_{k|k}(a_k,x_k) = \sum_{j=1}^{J^{1}_{k|k-1}} \omega^{1,j}_{\bar{D},k|k} \, \mathcal{N} (m^{1,j}_{k|k-1}, P^{1,j}_{k|k-1}) 
	+ \displaystyle \sum_{y_k \in \Z_k} \sum_{j=1}^{J^{1}_{k|k-1}} \omega^{1,j}_{D,k|k} \, \mathcal{N} (m^{1,j}_{k|k}, P^{1,j}_{k|k})
	\label{c:1+z} 
\end{eqnarray}
with weights 
%given by \eqref{wab1}.
\begin{eqnarray} 
	\omega^{1,j}_{\bar{D},k|k} &=& \frac{(1-p_d) \, \omega^{1,j}_{k|k-1}}{1-p_d+p_d \displaystyle \sum_{y_k \in \Z_k} \sum_{h=1}^{J^{1}_{k|k-1}} \frac{\omega^{1,h}_{k|k-1}}{n \, \kappa(y_k)} q^{1,h}_k (y_k)} \nonumber  \\
	\omega^{1,j}_{D,k|k} &=& \frac{p_d \, \omega^{1,j}_{k|k-1} q^{1,j}_k (y_k)}{\Big[ 1-p_d+p_d \displaystyle \sum_{y_k \in \Z_k} \sum_{h=1}^{J^{1}_{k|k-1}} \frac{\omega^{1,h}_{k|k-1}}{n \, \kappa(y_k)} q^{1,h}_k (y_k) \Big]  \, n \, \kappa(y_k)}   .  \nonumber
\end{eqnarray}

%\vspace{.3 cm}
\subsection{GM-HBF prediction}\label{sec5.3}

\begin{proposition} 
	Suppose assumptions \eqref{eq:GM_like}-\eqref{eq:GM_attack} hold, the posterior FISST density at time $k$ is fully specified by the triplet 
	$\big( r_{k|k}, p^0_{k|k}(x_k), p^1_{k|k}(a_k,x_k) \big)$,
	and $p^0_{k|k}(\cdot)$, $p^1_{k|k}(\cdot,\cdot)$ are Gaussian mixtures of the form \eqref{p0kk}-\eqref{p1kk}.
	%	\begin{eqnarray} 
	%		p^{0,i}_{k|k}(x_{k}) && \hspace{-.6cm} = \sum_{j=1}^{J^{0,i}_{k|k}} \omega^{0,ij}_{k|k} \mathcal{N} (\hat{x}^{0,ij}_{k|k}, P^{0,ij}_{k|k}) \label{inizpred0}
	%		\\
	%		p^{1,i}_{k|k}(a_k,x_k) &=& \sum_{j=1}^{J^{1,i}_{k|k}} \omega^{1,ij}_{k|k} \mathcal{N} (\tilde{x}^{1,ij}_{k|k}, \tilde{P}^{1,ij}_{k|k})
	%		\label{inizpred1} 
	%	\end{eqnarray} 
	Then the predicted FISST density 
	$\big( r_{k+1|k}, p^0_{k+1|k}(x_{k+1}), p^1_{k+1|k}(a_{k+1},x_{k+1}) \big)$
	is given by
	\begin{eqnarray} 
		\hspace{-.8cm}	r_{k+1|k} &=& (1 - r_{k|k}) \, p_b + r_{k|k} \, p_s 
		\\
		\hspace{-.8cm}	p^{0}_{k+1|k}(x_{k+1}) &=& \sum_{j=1}^{J^{0}_{k+1|k}} \omega^{0,j}_{k+1|k} \mathcal{N} (m^{0,j}_{k+1|k}, P^{0,j}_{k+1|k}) \label{p0pred}
		\\
		\hspace{-.8cm}	p^{1}_{k+1|k}(a_{k+1},x_{k+1}) &=& \sum_{j=1}^{J^{1}_{k+1|k}} \omega^{1,j}_{k+1|k} \mathcal{N} (m^{1,j}_{k+1|k}, P^{1,j}_{k+1|k})   \label{p1pred}
	\end{eqnarray} 
	where \eqref{p0pred} 
	%	can be written as 
	comprises $J^{0}_{k+1|k} = J^{0}_{k|k} + J^{1}_{k|k}$ components, i.e.
	\begin{eqnarray} 
		p^{0}_{k+1|k}(x_{k+1}) = 
		\underbrace{\sum_{j=1}^{J^{0}_{k|k}}  \omega^{0,j}_{\bar{B},k+1|k} \, \mathcal{N} (m^{0,j}_{\bar{B},k+1|k}, P^{0,j}_{\bar{B},k+1|k}) }_{\text{no attack-birth}} 
		+  \underbrace{\sum_{j=1}^{J^{1}_{k|k}} \omega^{0,j}_{\bar{S},k+1|k} \, \mathcal{N} (m^{0,j}_{\bar{S},k+1|k}, P^{0,j}_{\bar{S},k+1|k})   }_{\text{no attack-survival}} 
		\label{nbs}
	\end{eqnarray} 
	with 
	\begin{eqnarray} 
		m^{0,j}_{\bar{B},k+1|k} &=& 
		%		\hat{x}^{0,j}_{\bar{B},k+1|k} 
		%		=  
		A \, m^{0,j}_{k|k}  
		\label{mb}
		\\
		P^{0,j}_{\bar{B},k+1|k} &=& A P^{0,j}_{k|k} A^T + Q \label{Pb}
		\\
		\omega^{0,j}_{\bar{B},k+1|k} &=& \frac{(1 - r_{k|k}) (1 - p_b)}{1 - r_{k+1|k}} \, \omega^{0,j}_{k|k} 
		\label{wb}
	\end{eqnarray}
	and 
	\begin{eqnarray} 
		m^{0,j}_{\bar{S},k+1|k} &=& 
		%		\hat{x}^{h,j}_{\bar{S},k+1|k} 
		%		=  
		\tilde{A} \, m^{1,j}_{k|k}
		\label{ms}
		\\
		P^{0,j}_{\bar{S},k+1|k} &=& \tilde{A} P^{1,j}_{k|k} \tilde{A}^T + Q \label{Ps}
		\\
		\omega^{0,j}_{\bar{S},k+1|k} &=& \frac{ r_{k|k} \, (1 - p_s)}{1 - r_{k+1|k}} \, \omega^{1,j}_{k|k}   
		\label{ws}
	\end{eqnarray}
	where $\tilde{A} \defi [A, G]$.
	%		with weights
	%		\begin{eqnarray} 
	%			\omega^{h,j}_{\bar{B},k+1|k} &=& \frac{(1 - r_{k|k}) (1 - p_b)}{1 - r_{k+1|k}} \, \pi_{hi} \, \omega^{0,hj}_{k|k}  \\
	%			\omega^{h,j}_{\bar{S},k+1|k} &=& \frac{(1 - r_{k|k}) (1 - p_s)}{1 - r_{k+1|k}} \, \pi_{hi} \, \omega^{1,hj}_{k|k} 
	%		\end{eqnarray}
	Moreover, \eqref{p1pred} 
	comprises $J^{1}_{k+1|k} = J^{a} (J^{0}_{k|k} + J^{1}_{k|k})$ components, i.e.
	\small
	\begin{eqnarray} 
		p^{1}_{k+1|k}(a_{k+1},x_{k+1})   
		= \underbrace{ \sum_{j=1}^{J^{0}_{k|k}}  \sum_{h=1}^{J^{a}} \omega^{1,jh}_{B,k+1|k} \, \mathcal{N} (m^{1,jh}_{B,k+1|k}, P^{1,jh}_{B,k+1|k})  }_{\text{attack-birth}} 
		+ \underbrace{  \sum_{j=1}^{J^{1}_{k|k}} \sum_{h=1}^{J^{a}} \omega^{1,jh}_{S,k+1|k} \, \mathcal{N} (m^{1,jh}_{S,k+1|k}, P^{1,jh}_{S,k+1|k}) }_{\text{attack-survival}}
		\label{bs}
	\end{eqnarray}
 \normalsize
	%%%&& \hspace{-4cm}  = \underbrace{ \sum_{i=1}^{J^B_{k+1|k}} \omega^{B,i}_{k+1|k} \, \mathcal{N} (\hat{x}^{B,i}_{k+1|k}, P^{B,i}_{k+1|k}) }_{\text{attack-birth}}
	%%%+ \underbrace{  \sum_{j=1}^{J^S_{k+1|k}} \omega^{S,j}_{k+1|k} \, \mathcal{N} (\hat{x}^{S,j}_{k+1|k}, P^{S,j}_{k+1|k})    }_{\text{attack-survival}}
	where
	\begin{eqnarray} 
		m^{1,jh}_{B,k+1|k} &=& 
		\begin{bmatrix} 
			A \, m^{0,j}_{k|k}  \\
			\tilde{a}^{h}
		\end{bmatrix}
		\\
		P^{1,jh}_{B,k+1|k} &=&  \begin{bmatrix} 
			A P^{0,j}_{k|k} A^T + Q & 0 \\
			0 & \tilde{P}^{a,h} \end{bmatrix} \\
		\omega^{1,jh}_{B,k+1|k} &=& \frac{(1 - r_{k|k}) \, p_b}{r_{k+1|k}} \, \omega^{0,j}_{k|k} \, \tilde{\omega}^{a,h}
		\label{abirth}
	\end{eqnarray}
	and 
	\begin{eqnarray} 
		m^{1,jh}_{S,k+1|k} &=&
		\begin{bmatrix} 
			\tilde{A} \, m^{1,j}_{k|k} \\
			\tilde{a}^{h}
		\end{bmatrix}
		%			A \, \hat{x}^{1,j}_{k|k} + G \, \hat{a}^{h}_{k+1}
		\\
		%m^{h,jl}_{S,k+1|k} &=& A^h \hat{x}^{1,hj}_{k|k} + G^h \hat{a}^{hl}_{k+1}   \\
		P^{1,jh}_{S,k+1|k} &=& \begin{bmatrix} 
			\tilde{A} P^{1,j}_{k|k} \tilde{A}^T + Q & 0 \\
			0 & \tilde{P}^{a,h} \end{bmatrix} 
		\\
		\omega^{1,jh}_{S,k+1|k} &=& \frac{r_{k|k} \, p_s}{r_{k+1|k}}  \, \omega^{1,j}_{k|k} \, \tilde{\omega}^{a,h}  .
		\label{asurv}
	\end{eqnarray}
	%		with weights
	%		\begin{eqnarray} 
	%			\omega^{h,jl}_{B,k+1|k} &=& \frac{(1 - r_{k|k}) \, p_b}{r_{k+1|k}} \, \pi_{hi} \, \omega^{0,hj}_{k|k} \, \omega^{a,hl}_{k+1} \\
	%			\omega^{h,jl}_{S,k+1|k} &=& \frac{r_{k|k} \, p_s}{r_{k+1|k}} \, \pi_{hi} \, \omega^{1,hj}_{k|k} \, \omega^{a,hl}_{k+1}
	%		\end{eqnarray}	
\end{proposition} 
%\vspace{.3 cm}
%{\em Proof:}

\textit{Proof:}
The predicted signal attack probability comes directly from \eqref{p_existence1}. Let us now derive the predicted density $p^{0}_{k+1|k} (\cdot)$.
%then $p^{0,i}_{k|k}$ can be obtained analogously.
From \eqref{p_01} in Theorem 3:   
\begin{eqnarray} 
	p^{0}_{k+1|k}(x_{k+1}) 
	&=& \frac{(1 - r_{k|k}) \, (1 - p_b) }{1 - r_{k+1|k}}  
	\int \pi(x_{k+1}|x_{k}) , p^0_{k|k}(x_{k}) \, \mbox{d} x_{k} \nonumber \\
&& + \frac{r_{k|k} \, (1-p_s) }{1 - r_{k+1|k}} 
	\iint \pi(x_{k+1}|a_{k},x_{k}), p^1_{k|k}(a_{k},x_{k}) \, \mbox{d}a_k \mbox{d}x_{k}  .
\end{eqnarray} 
Using \eqref{eq:GM_trans}, \eqref{p0kk} in the first term and \eqref{eq:GM_trans2}, \eqref{p1kk} in the second term,
%, and recalling the definition of transitional probabilities \eqref{transprob}
we can  rewrite   
\begin{eqnarray}  
	p^{0}_{k+1|k}(x_{k+1}) 
	&=& \frac{(1 - r_{k|k}) \, (1 - p_b)}{1 - r_{k+1|k}}
	\int \mathcal{N}  (x; A x_k, Q) 
	\sum_{j=1}^{J^{0}_{k|k}} \omega^{0,j}_{k|k} \, \mathcal{N} (m^{0,j}_{k|k}, P^{0,j}_{k|k})  \, \mbox{d} x_{k}  \nonumber \\
	&& + \frac{r_{k|k} \, (1-p_s)}{1 - r_{k+1|k}}  
\iint \mathcal{N}  (x; A x_k + G a_k, Q) 
	\sum_{j=1}^{J^{1}_{k|k}} \omega^{1,j}_{k|k} \, \mathcal{N} (m^{1,j}_{k|k}, P^{1,j}_{k|k})  \, \mbox{d} a_{k} \mbox{d} x_{k} .
%	\nonumber 
\end{eqnarray} 
Hence, using Lemma 1 by Vo and Ma \cite{Vo2006} in both the above terms, we finally derive \eqref{nbs}:
\begin{eqnarray} 
p^{0}_{k+1|k}(x_{k+1}) 
	&=& \sum_{j=1}^{J^{0}_{k|k}}  
	\frac{(1 - r_{k|k}) \, (1 - p_b)}{1 - r_{k+1|k}} \, \omega^{0,j}_{k|k} \,
	\mathcal{N}  (x; A m^{0,j}_{k|k}, A P^{0,j}_{k|k} A^T + Q) 
	\nonumber \\
&&	+ \sum_{j=1}^{J^{1}_{k|k}}
	\frac{r_{k|k} \, (1-p_s)}{1 - r_{k+1|k}} 
	\, \omega^{1,j}_{k|k} \,  
	\mathcal{N}  (x; \tilde{A} m^{1,j}_{k|k}, \tilde{A} P^{1,j}_{k|k} \tilde{A}^T + Q) .  \nonumber
\end{eqnarray} 
%\small{
%	\begin{eqnarray} 
%	&&\hspace{-.5cm}p^{1,i}_{k+1|k}(a_{k+1},x_{k+1}) \\
%	&&\hspace{-.5cm}=\underbrace{  \sum_{h \in \mathcal{M}} \sum_{j=1}^{J^{0,h}_{k|k}}  \sum_{l=1}^{J^{a,h}_{k+1}} \frac{(1 - r_{k|k}) \, p_b}{r_{k+1|k}} \, \pi_{hi} \, \omega^{0,hj}_{k|k}  \omega^{a,hl}_{k+1}
%		\mathcal{N} (m^{h,jl}_{B,k+1|k}, P^{h,jl}_{B,k+1|k})   }_{\text{attack-birth}} \nonumber
%	\\
%	&&\hspace{-.5cm} +  \underbrace{  \sum_{h \in \mathcal{M}}   \sum_{j=1}^{J^{1,h}_{k|k}} \sum_{l=1}^{J^{a,h}_{k+1}} \frac{r_{k|k} \, p_s}{r_{k+1|k}} \pi_{hi} \, \omega^{1,hj}_{k|k} \omega^{a,hl}_{k+1} \, \mathcal{N} (m^{h,jl}_{S,k+1|k}, P^{h,jl}_{S,k+1|k}) }_{\text{attack-survival}}  \nonumber  
%	%%%&& \hspace{-4cm}  = \underbrace{ \sum_{i=1}^{J^B_{k+1|k}} \omega^{B,i}_{k+1|k} \, \mathcal{N} (\hat{x}^{B,i}_{k+1|k}, P^{B,i}_{k+1|k}) }_{\text{attack-birth}}
%	%%%+ \underbrace{  \sum_{j=1}^{J^S_{k+1|k}} \omega^{S,j}_{k+1|k} \, \mathcal{N} (\hat{x}^{S,j}_{k+1|k}, P^{S,j}_{k+1|k})    }_{\text{attack-survival}}
%	\end{eqnarray} } \normalsize
%which can be finally expressed as in \eqref{bs}.
In a similar fashion, we can obtain $p^{1}_{k+1|k}(\cdot,\cdot)$. From \eqref{p_11} in Theorem 3:
%p^0_{k+1|k}(x_{k+1},\nu_{k+1}) =&& \hspace{-.6cm} \frac{(1 - r_{k|k}) (1 - p_b) \, p_{k+1|k}(x_{k+1}|\emptyset)}{1 - r_{k+1|k}}
\begin{eqnarray} 
p^{1}_{k+1|k}(a_{k+1},x_{k+1})
	&=& \frac{(1 - r_{k|k}) \, p_b }{r_{k+1|k}}  
	\int \pi(x_{k+1}|x_{k}) , p^0_{k|k}(x_{k}) \, \mbox{d} x_{k} \, p(a) 
		\nonumber \\
&& + \frac{r_{k|k} \, p_s }{r_{k+1|k}} 
	\iint \pi(x_{k+1}|a_{k},x_{k}) \, p^1_{k|k}(a_{k},x_{k}) \, \mbox{d}a_k \mbox{d}x_{k} \, p(a) 
	\nonumber
\end{eqnarray}  
which, using \eqref{eq:GM_trans}, \eqref{eq:GM_trans2}, \eqref{eq:GM_attack}, \eqref{p0kk} and \eqref{p1kk}, leads to
%	\begin{eqnarray} 
%		&& \hspace{-.5cm}p^{0,i}_{k+1|k}(x_{k+1}) 
%		\\
%		&&\hspace{-.5cm} =\frac{(1 - r_{k|k}) (1 - p_b)}{1 - r_{k+1|k}} 
%		\sum_{h \in \mathcal{M}} \pi_{hi}
%		\int \mathcal{N}  (x; A^h x_k, Q^h)   \nonumber\\
%		&& \hspace{-.5cm}\sum_{j=1}^{J^{0,h}_{k|k}} \omega^{0,hj}_{k|k} \, \mathcal{N} (m^{0,hj}_{k|k}, P^{0,hj}_{k|k}) \, \mbox{d} x_{k} \nonumber  \\
%		&&  \hspace{-.5cm}+ \frac{r_{k|k} (1 - p_s)}{1 - r_{k+1|k}} 
%		\sum_{h \in \mathcal{M}} \pi_{hi}
%		\iint \mathcal{N}  (x; A^h x_k + G^h a_k, Q^h) \nonumber \\
%		&&\hspace{-.5cm} \sum_{j=1}^{J^{1,h}_{k|k}} \omega^{1,hj}_{k|k} \, \mathcal{N} (m^{1,hj}_{k|k}, P^{1,hj}_{k|k})  
%		\sum_{j=1}^{J^{1,h}_{k|k}} \omega^{1,hj}_{k|k} \, \mathcal{N} (m^{1,hj}_{k|k}, P^{1,hj}_{k|k}) \, \mbox{d} x_{k}    \nonumber
%\end{eqnarray} 
\small
\begin{eqnarray} 
p^{1}_{k+1|k}(a_{k+1},x_{k+1}) 
	&=& \frac{(1 - r_{k|k}) \, p_b}{r_{k+1|k}} 
	\int \mathcal{N}  (x; A x_k, Q) 
\sum_{j=1}^{J^{0}_{k|k}} \omega^{0,j}_{k|k} \, \mathcal{N} (m^{0,j}_{k|k}, P^{0,j}_{k|k})  \, \mbox{d} x_{k}
	\sum_{h=1}^{J^{a}} \tilde{\omega}^{a,h} \, \mathcal{N} (a;  \tilde{a}^{h},\tilde{P}^{a,h})   \nonumber  \\
	&& \hspace{-2.5cm} + \frac{r_{k|k} \, p_s}{r_{k+1|k}} 
	\iint \mathcal{N}  (x; A x_k + G a_k, Q) 
\sum_{j=1}^{J^{1}_{k|k}} \omega^{1,j}_{k|k} \, \mathcal{N} (m^{1,j}_{k|k}, P^{1,j}_{k|k})  \, \mbox{d} a_{k} \mbox{d} x_{k}  
	\sum_{h=1}^{J^{a}} \tilde{\omega}^{a,h} \, \mathcal{N} (a;  \tilde{a}^{h},\tilde{P}^{a,h}) . 
\end{eqnarray} 
\normalsize
Finally, by applying the same result on integrals of Gaussians used above, we obtain \eqref{bs}: 
\small
\begin{eqnarray} 
   p^{1}_{k+1|k}(a_{k+1},x_{k+1}) 
	&=& \sum_{j=1}^{J^{0}_{k|k}} \sum_{h=1}^{J^{a}}
	\frac{(1 - r_{k|k}) \, p_b}{r_{k+1|k}} 
	\, \omega^{0,j}_{k|k} \, \tilde{\omega}^{a,h} 
\mathcal{N}  (x; A m^{0,j}_{k|k}, A P^{0,j}_{k|k} A^T + Q) 
	\, \mathcal{N} (a;  \tilde{a}^{h},\tilde{P}^{a,h})   \nonumber  \\
	&&  + 
	\sum_{j=1}^{J^{1}_{k|k}} \sum_{h=1}^{J^{a}}
	\frac{r_{k|k} \, p_s}{r_{k+1|k}} 
	\, \omega^{1,j}_{k|k} \,  \tilde{\omega}^{a,h} 
\mathcal{N}  (x; \tilde{A} m^{1,j}_{k|k}, \tilde{A} P^{1,j}_{k|k} \tilde{A}^T + Q) 
	\, \mathcal{N} (a;  \tilde{a}^{h},\tilde{P}^{a,h}) .   
\end{eqnarray} 
\normalsize

%\vspace{.3cm}
It is worth pointing out that, likewise other GM filters, 
%(e.g. \cite{Vo2006}), 
also the proposed \textit{Gaussian Mixture Hybrid Bernoulli Filter}
%turns out to be affected by computational issues as
is characterized by
a number of Gaussian components that increases with no bound over time. 
As already noticed in the above derivation, at time $k$ the GM-HBF requires 
%\begin{eqnarray}
%	J^0_{k|k} &=& \left\{ \ba{ll} 
%	2 \, J^0_{k|k-1},                  & \mbox{packet substitution }     \vspace{2mm} \nonumber  \\
%	J^0_{k|k-1} (1 + |\Z_k|),                & \mbox{extra packet injection}                                 
%	\ea \right.   \nonumber  \\
%	J^1_{k|k} &=& \left\{ \ba{ll} 
%	2 \, J^1_{k|k-1},                  & \mbox{packet substitution }     \vspace{2mm} \nonumber    \\
%	J^1_{k|k-1} (1 + |\Z_k|),                & \mbox{extra packet injection}               \nonumber                  
%	\ea \right.
%\end{eqnarray}
\small
\begin{eqnarray}
J^0_{k|k} = \left\{ \ba{ll} 
2 \, J^0_{k|k-1},                  & \mbox{packet substitution }     \vspace{2mm} \nonumber  \\
J^0_{k|k-1} (1 + |\Z_k|),                & \mbox{extra packet injection}                                 
\ea \right.,   \quad J^1_{k|k} &=& \left\{ \ba{ll} 
2 \, J^1_{k|k-1},                  & \mbox{packet substitution }     \vspace{2mm} \nonumber    \\
J^1_{k|k-1} (1 + |\Z_k|),                & \mbox{extra packet injection}               \nonumber                  
\ea \right.
\end{eqnarray} \normalsize
components to exactly represent the posterior densities $p^0_{k|k}(\cdot)$ and $p^1_{k|k}(\cdot,\cdot)$, respectively. 
Here 
\begin{eqnarray} 
	J^0_{k|k-1} &=& J^0_{k-1|k-1} + J^1_{k-1|k-1}, \nonumber \\
	J^1_{k|k-1} &=& J^a (J^0_{k-1|k-1} + J^1_{k-1|k-1}) \nonumber
\end{eqnarray} 
denote the number of components generated in the prediction step.
Heuristic \textit{pruning} and \textit{merging} procedures \cite{Vo2006} can be performed at each time step so as to remove low-weight components and combine statistically close components and, hence, reduce the growing number of GM components.

\begin{remark}
The Gaussian-mixture implementation of this section has actually revealed a connection between the proposed hybrid Bernoulli filter and the Kalman filter (KF) in that the former uses multiple KFs
(or EKFs/UKFs) to propagate in time means and covariances of the various components of the Gaussian mixture
(see eqns. (\ref{cgmx0_ps})-(\ref{S0}), (\ref{cgmx1_ps})-(\ref{cPxa_ps}), (\ref{mb})-(\ref{Pb}) and (\ref{ms})-(\ref{Ps})).
\end{remark}

% % % % % % % % % % % % % % % % % % % % % % % % % % % % % % %

\section{Numerical examples}\label{sec6}

The effectiveness of the developed tools, based on Bayesian random-set theory, for joint attack detection and secure state estimation of cyber-physical systems has been tested on two numerical examples concerning a benchmark linear dynamical system and a standard IEEE power network case-study.
Simulations have been carried out in the presence of both signal and extra packet injection attacks as well as uncertainty on measurement delivery.
Results on the performance of the GM-HBF under packet substitution attack are shown in Section~\ref{sec6.2}.

\subsection{Benchmark linear system}\label{sec6.1}

\begin{figure}[t!] 
	\centering
	\includegraphics[width=.5\columnwidth]{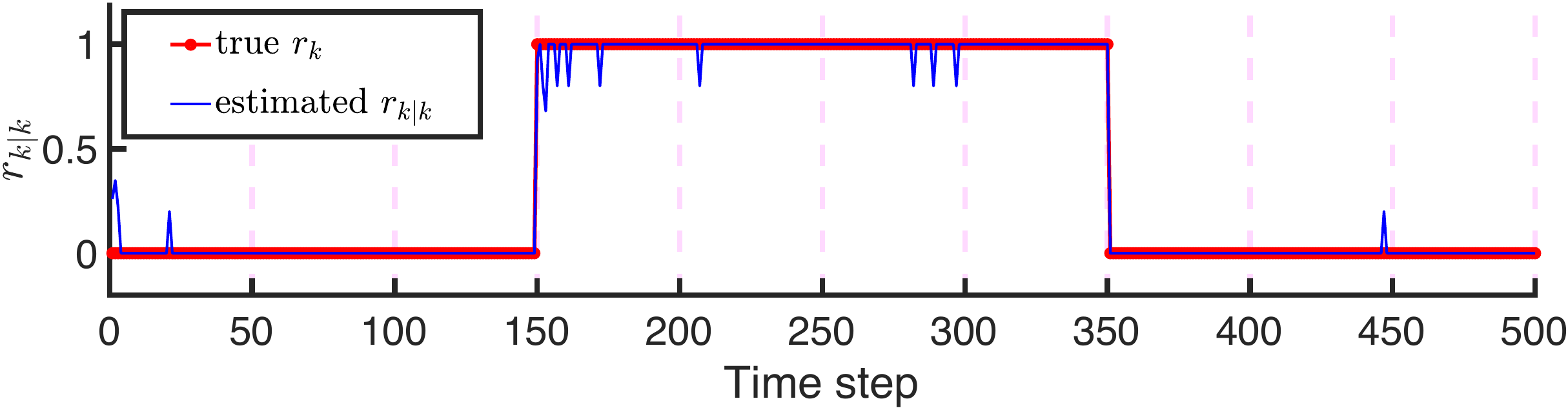}
	\caption{True and estimated attack probability.}
	\label{fig:exis}
\end{figure}

\begin{figure}[t!]
	\centering
	\includegraphics[width=.5\columnwidth]{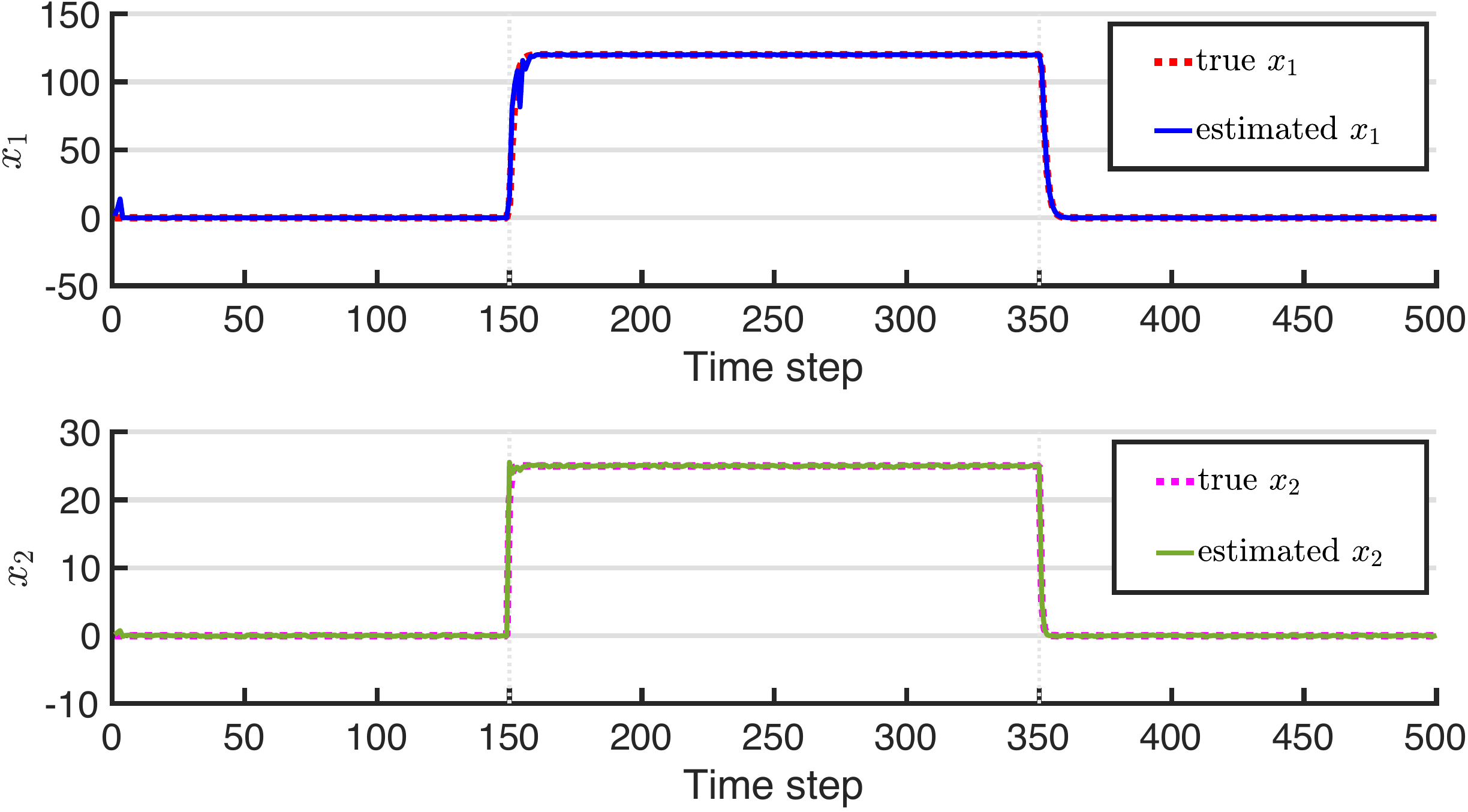}
	\caption{True and estimated state components $x_1$ and $x_2$.}
	\label{fig:x12}
\end{figure}

\begin{figure}[t!]
	\centering
	\includegraphics[width=.5\columnwidth]{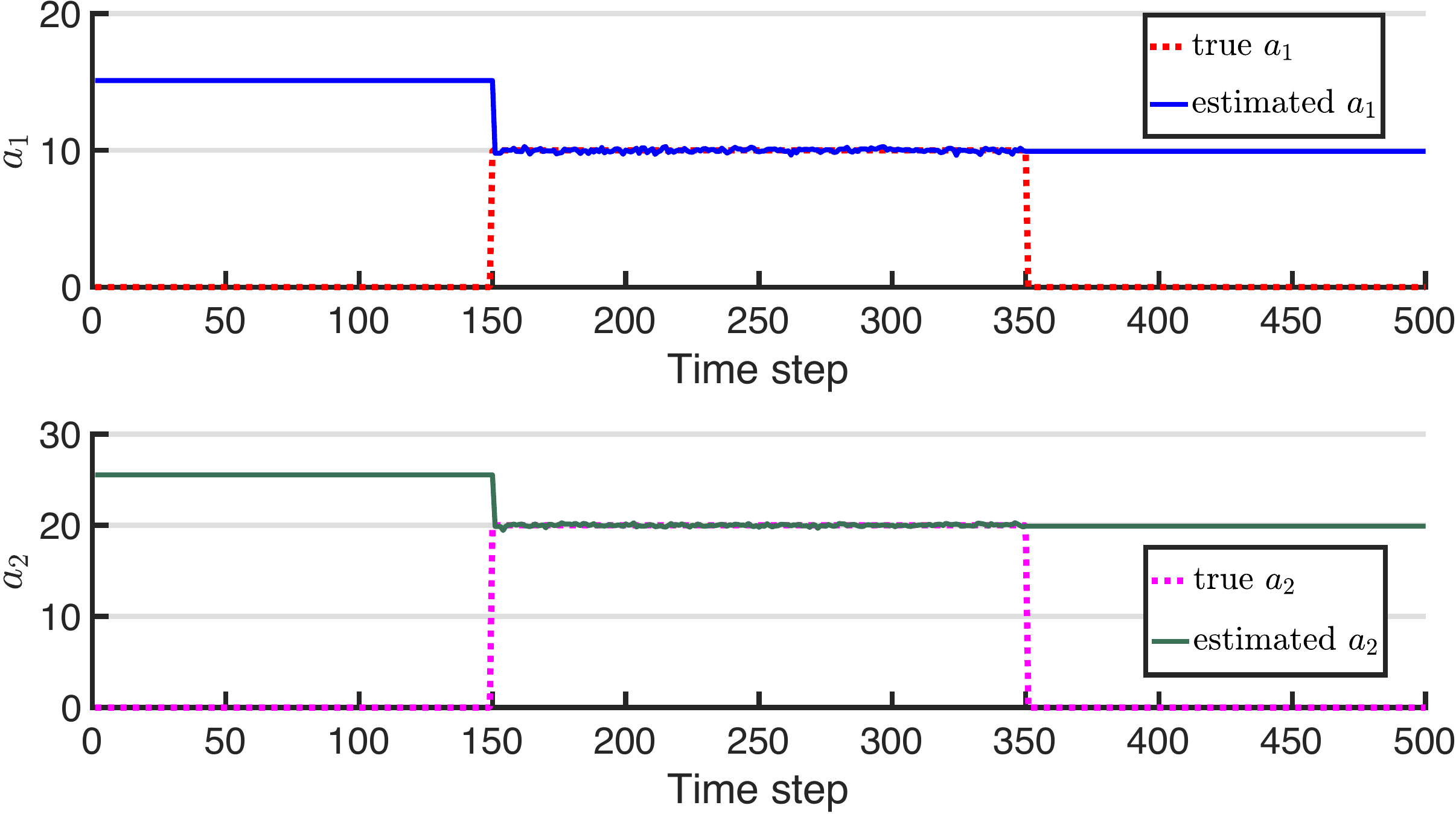}
	\caption{True and estimated attack components $a_1$ and $a_2$.}
	\label{fig:a12}
\end{figure}

Let us first consider the following benchmark linear system, already used in the JISE literature \cite{Cheng2009}: 
%obtained from the example in \cite{Yong2015}:
\begin{equation}\label{eq:linear_system}
\begin{array}{rcl}
x_{k+1} &=& A x_k + G a_k + w_k\\
y_k &=& C x_k + H a_k + v_k
\end{array}
\end{equation}
where $A$, $C$, $R$, and $Q$ are the same as in Yong et al. \cite{Yong2015},  while 
$G = [e_1,e_2]$ and $H = [e_3,e_1]$, where $e_1,\dots,e_5$ denote the canonical basis vectors.
For this numerical study, the probabilities of attack-birth and attack-survival are fixed, respectively, at $p_b = 0.2$ and $p_s = 0.8$. The system-generated measurement is supposed to be delivered at the monitor/control center with probability $p_d = 0.98$, while the initial signal attack probability is set to $r_{1|0} = 0.1$.
The initial state has been set equal to $x_0 = 0$, whereas both densities $p^0(\cdot)$ and $p^1(\cdot,\cdot)$ have been initialized as single Gaussian components with first guess mean
$\hat{x}_{1|0}^0 = [10,10,0,0,0]^T$
and covariance $P_{1|0}^0 = 10^4 \, I_5$. Moreover, the first estimate of the attack vector has been randomly initialized as $\tilde{a}_{1|0} = [15.1,25.53]^T$, with associated initial covariance matrix $\tilde{P}^a_{1|0} = 50 \, I_2$.
%, and $P^{ax}_{1|0} = 0$.
The extra fake measurements are modeled as uniformly distributed over the interval $[-0.3,140.3]$. 
Finally, a pruning threshold $\gamma_p = 10^{-3}$ and a merging threshold $\gamma_m = 3$ have been chosen. 
As shown in Fig.\ref{fig:exis}, at time $k=150$ a signal attack vector $a = [10,20]^T$ is injected into the system, persisting for $200$ time steps. The proposed GM-HBF promptly detects the unknown signal attack, by simply comparing the attack probability $r_{k|k}$ obtained in (\ref*{p_existence3}) with the threshold $0.5$.
Fig. \ref{fig:x12} provides a comparison between the true and the estimated values of states $x_1$ and $x_2$ (clearly the only state components affected by the signal attack). Note that the state estimate is obtained by means of a MAP estimator, i.e. by extracting the Gaussian mean with the highest weight from the posterior density $p^0(\cdot)$ (\ref{p_03}) or $p^1(\cdot,\cdot)$ (\ref{p_13}), according to the current value of the attack probability.
Finally, Fig. \ref{fig:a12} shows how the attack estimates extracted from $p(a)$ of the two components of the attack vector, coincide with the actual values inside the attack time interval $[150,350]$. Note that outside that interval the estimates of the attack vector are not meaningful because the attack probability $r_{k|k}$ is almost $0$.
	%\vspace{-.3cm}
%\vspace{-1.5cm}

% % % % % % % % % % % % % % % % % % % % % % % % % % % % % % % %

\subsection{IEEE 14-bus power network}\label{sec6.2}

\begin{figure}[tb]
	\begin{center}
		\includegraphics[width=.46\columnwidth]{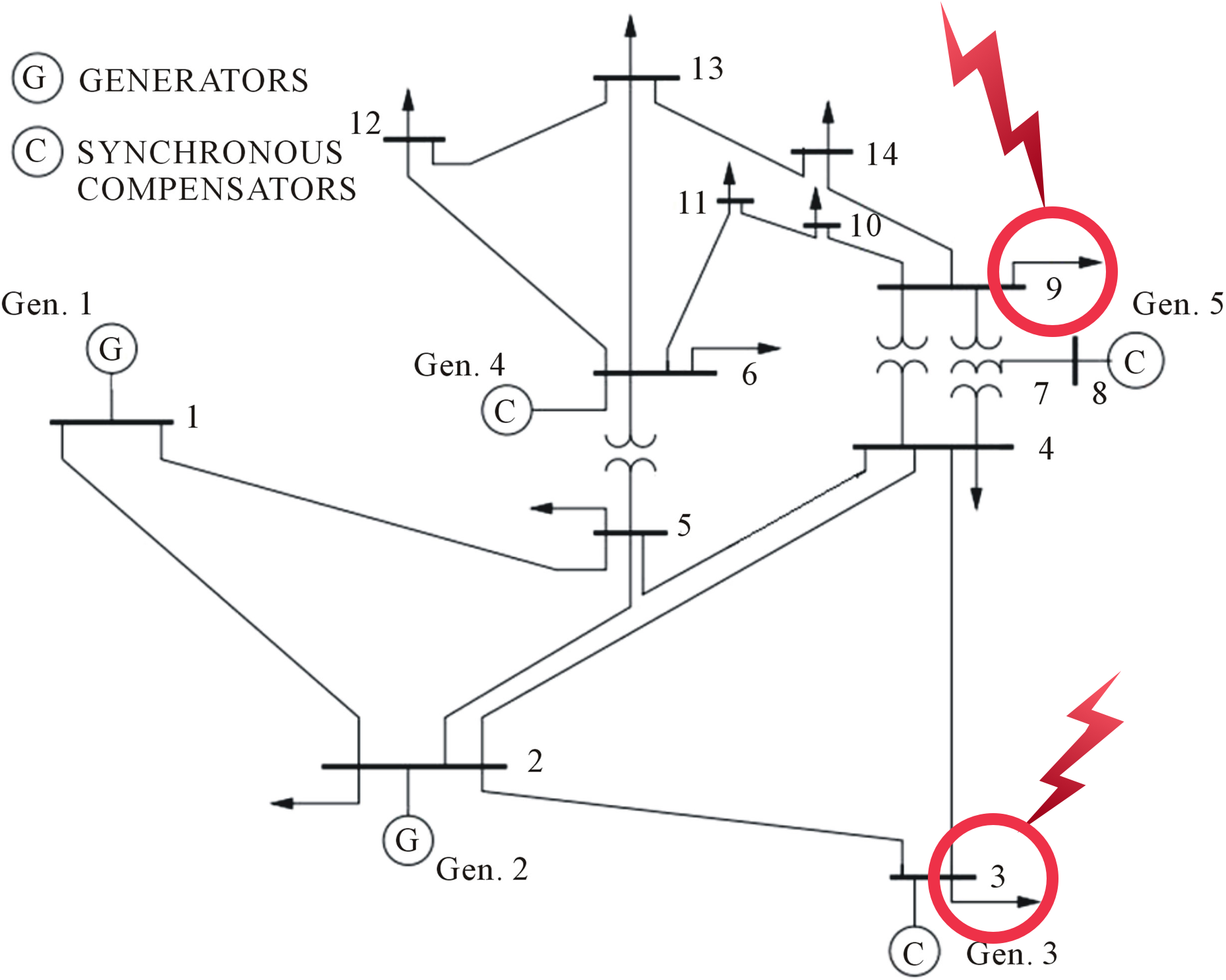}
		\caption{Single-line model of the IEEE 14-bus system. The \textit{true} victim load buses 3 and 9 are circled in red.}
		\label{fig:ieee14}
	\end{center}
\end{figure} 
\begin{figure}[tb]
	\begin{center}
		\includegraphics[width=.45\columnwidth]{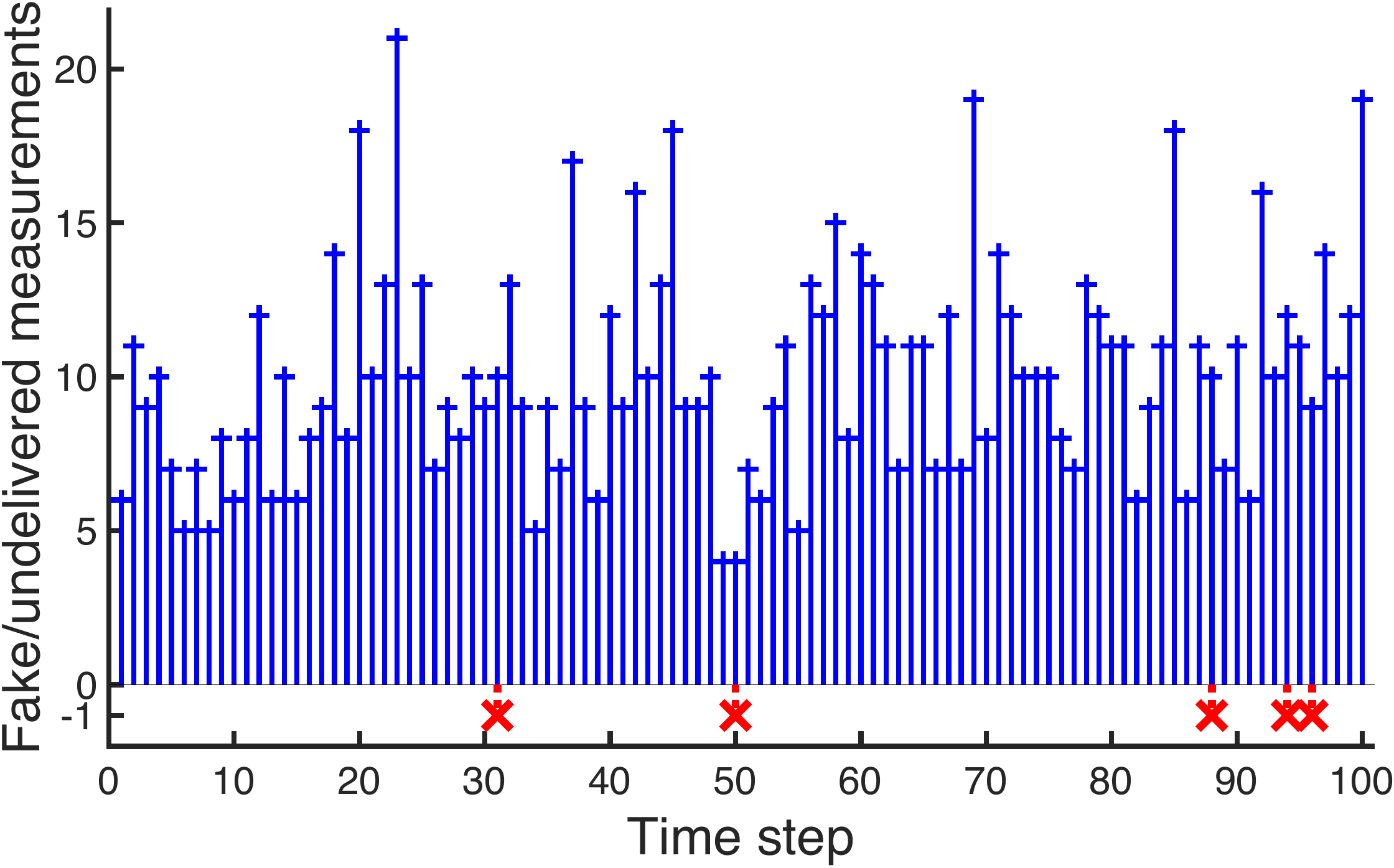}
		\caption{Number of extra fake measurements injected (blue circles) and undelivered ($p_d = 0.95$) system-originated observations (red cross in $-1$) vs time.
			The proposed GM-HB filter turns out to be particularly robust to \textit{extra packet injection} attacks.}
		\label{fig:extra}
	\end{center}
\end{figure}
\begin{figure}[tb]
	\begin{center}
		\includegraphics[width=.73\columnwidth]{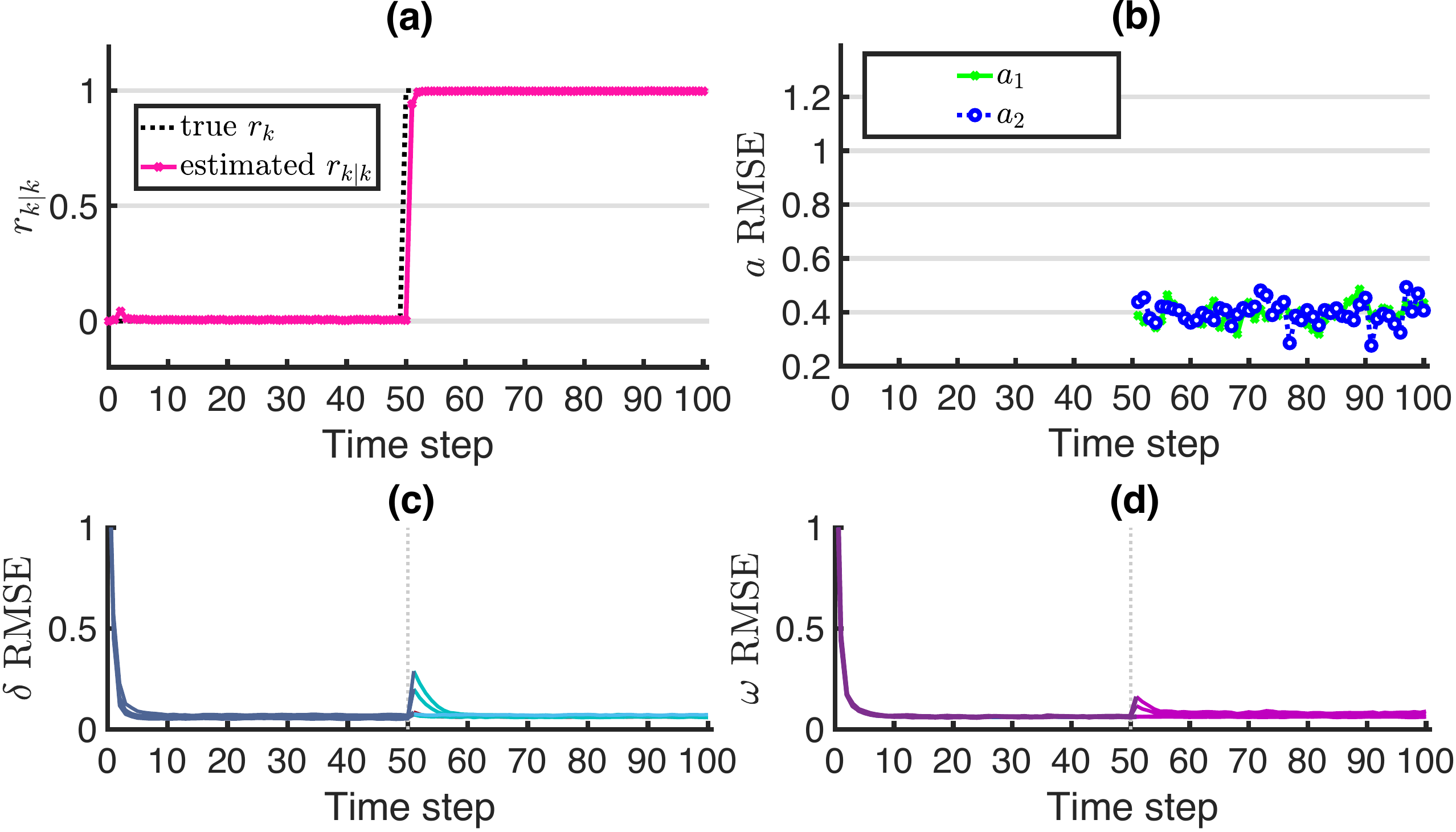}
		%		\caption{\textit{True} and estimated probability of existence $r_t^1$ of the signal attack $a_t$.}
		\caption{Performance of the GM-HBF in terms of joint attack detection (a) and estimation of attack signal (b), rotor angles $\delta_i, \, i=1,\dots,5$ (c), and frequencies $\omega_i, \, i=1,\dots,5$ (d).}
		\label{fig:rmse}
	\end{center}
\end{figure}

\begin{figure}[t!]
	\begin{center}
		\includegraphics[width=.55\columnwidth]{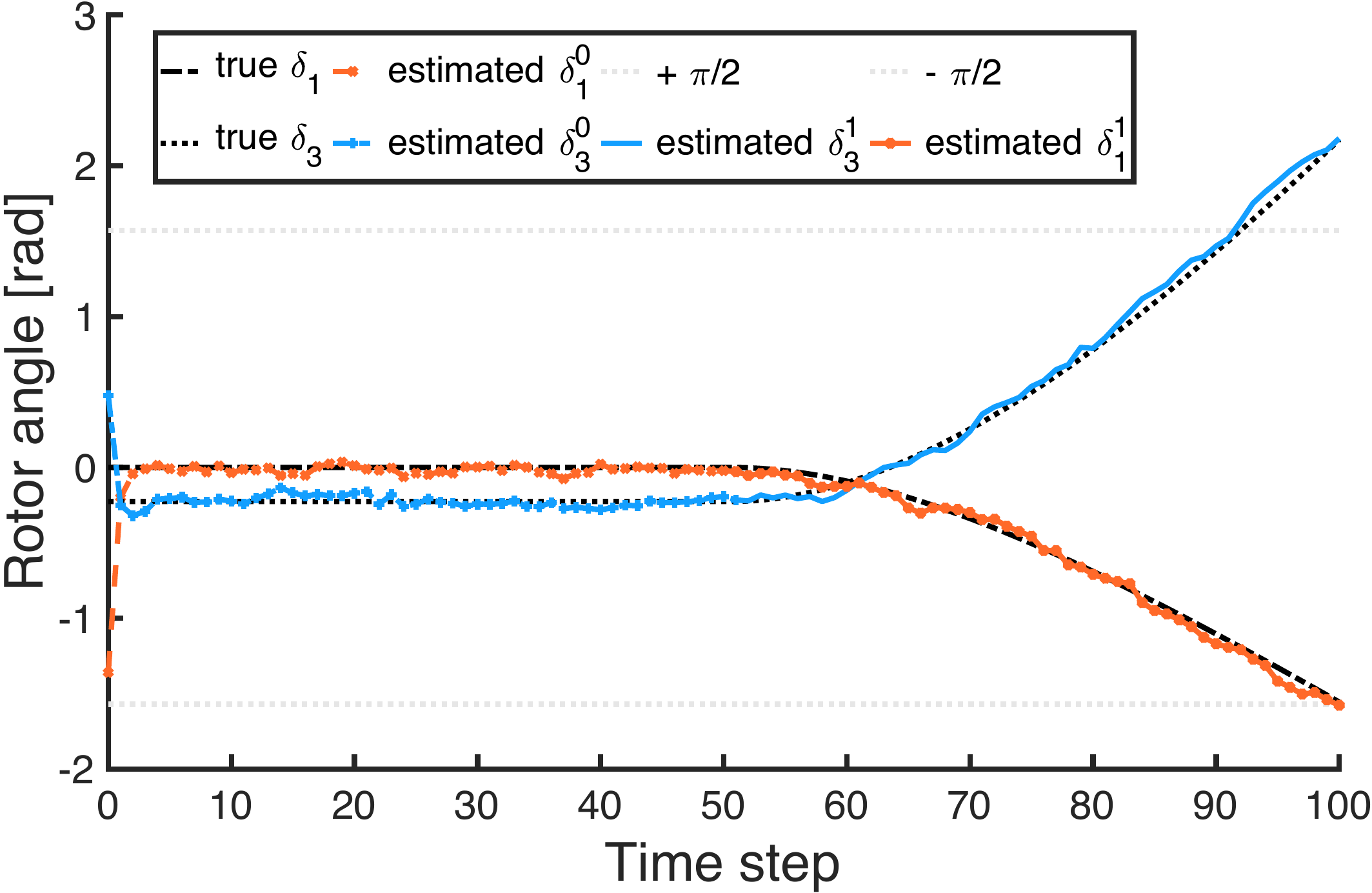}
		\caption{Estimated vs \textit{true} trajectory of rotor angles $\delta_j, \, j=1,3$. Note that, if $|\delta_j|$ is sufficiently large (values close to $\pi/2$), the linear small signal approximation significantly deviates from the nonlinear dynamics of the system, and hence the assumed dynamic model becomes inaccurate.}
		\label{fig:delta}
	\end{center}
\end{figure}

 
State estimation is of paramount importance to ensure the reliable operation of energy
delivery systems since it provides estimates of the power grid state by processing meter
measurements and exploiting power system models. Cyber attacks on power systems can
alter available information at the control center and generate fake meter and input data,
potentially causing power outage and forcing the energy management system to make
erroneous decisions, e.g. on contingency analysis and economic dispatch.
The proposed GM-HBF was tested on the IEEE 14-bus system (Fig.~\ref{fig:ieee14})
consisting of $5$ synchronous generators and $11$ load buses, with parameters taken from
%the 14-bus case of
%relative to transmission lines, generators' inertia and damping, the nominal power injections and demands are the same considered in the 14-bus case of
%\cite{SaPa:98}.
MATPOWER \cite{ZiMuTh:11}. 
The dynamics of the system can be described by the linearized swing equation
\cite{Kundur1994} 
%for the $n = 6$ active buses, 
derived through the Kron reduction \cite{PaBiBu11} of the linear small-signal power network model. 
The DC state estimation model assumes $1$ p.u. (per unit) voltage magnitudes in all buses and $j1$ p.u. branch impedance, with $j$ denoting imaginary unit.
The system dynamics is represented by the evolution of $n=10$ states comprising both the rotor angles $\delta_j$ and the frequencies $\omega_j$ of each generator $j$ in the network.
After discretization (with sampling interval $T = 0.01 s$), the model of the system takes the form \eqref{sys1}-\eqref{sys2},
where the whole state is measured by a network $\mathcal{S}_i$ of sensors. 
The system is assumed to be corrupted by additive zero mean Gaussian white process and measurement noises with variances $\sigma_w^2=0.01$ and $\sigma_{v}^2=0.01$.
At time $k=50$ a signal attack vector $a = [0.2,0.1]^T$ p.u. is injected into the system to abruptly increase the real power demand of the two victim load buses $3$ and $9$ with an additional loading of $21.23 \%$ and, respectively, $33.9 \%$. 
This type of attack, referred to as \textit{load altering attack} \cite{AmMoPa:15}, 
%	This in turn causes a loss of synchrony of the rotor angles (Fig. ) as we can notice from the large deviation of the rotor speeds of all generators from the nominal value $\omega_s = 60$ Hz (Fig.\ref{}).
can provoke a loss of synchrony of the rotor angles and hence a deviation of the rotor speeds of all generators from their nominal value.
%, here $\omega_s = 60$ Hz.
%In this test case, the probabilities of attack-birth and attack-survival are fixed, respectively, at $p_b = 0.05$ and $p_s = 0.95$, while
%the system-generated measurement vector is supposed to be delivered at the local fusion node with probability $p_d = 0.99$.
In addition, we fixed the following parameters: $p_b = 0.05$, $p_s = 0.95$, $p_d = 0.95$, pruning and merging thresholds $\gamma_p=10^{-2}$ and $\gamma_m=3$ for the Gaussian-mixture implementation. 
%    , while the initial signal attack probability is set to $r_{1|0} = 0.4$.
Let us first consider the system under extra packet injection attack.
The additional fake measurements injected into the sensor channels are modeled 
%as a Poisson RFS
%	uniformly distributed over the interval $[-0.3,140.3]$, 
%with average number $\xi = 10$ and 
as uniformly distributed over the interval $[-10,5]$, suitably chosen to emulate system-originated observations. 
Fake and missed packets are shown in Fig.~\ref{fig:extra} for a specific run.
The joint attack detection and state estimation performance of the GM-HBF algorithm has been analyzed by Monte Carlo simulations.
Fig.~\ref{fig:rmse} shows the \textit{true} and estimated probability of attack existence (a) and the Root Mean Square Error (RMSE), averaged over $1000$ Monte Carlo runs, relative to the rotor angle (b) and frequency (c) estimates. 
Fig.~\ref{fig:rmse} (d) shows the RMSE of the estimated components of the signal attack, extracted from $p_{k|k}^1(a,x)$. 
As shown in the results (a)-(d), the proposed secure state estimator succeeds in promptly detecting a signal attack altering the nominal energy delivery system behavior, 
and hence in being simultaneously resilient to integrity attacks on power demand, and robust to extra fake packets and undelivered measurements.
%Fig.~\ref{fig:extra} shows the resulting number of fake measurements maliciously injected at each time step. 
%Figs.~\ref{fig:probexist}-\ref{fig:attack} show the performance of the  filter. In particular, Fig.~\ref{fig:probexist} displays the true and estimated attack existence probability.
%\begin{figure}[h!]
%	\begin{center}
%		\includegraphics[width=\columnwidth]{./sub41_r_delta_omega_attack_RMSE_all}
%		\caption{Performance of the GM-HB filter in terms of joint attack detection (a) and estimation of rotor angles $\delta_i, \, i=1,\dots,5$ (b), frequencies $\omega_i, \, i=1,\dots,5$ (c), and attack signal (d).}
%		\label{fig:rmse}
%	\end{center}
%\end{figure} 
Fig.~\ref{fig:delta} provides, for a single Monte Carlo trial, a comparison between the \textit{true} and the estimated values of the two rotor angles mainly affected by the victim load buses, and clearly shows how $\delta_1$ and $\delta_3$ lose synchrony once the load altering attack enters into action. 
Nevertheless, the proposed secure filter keeps tracking the state evolution with high accuracy even after time $k=50$, once recognized that the system is under attack.
%\vspace{-1.3cm}
\begin{figure}[tb]
	\begin{center}
		\includegraphics[width=.54\columnwidth]{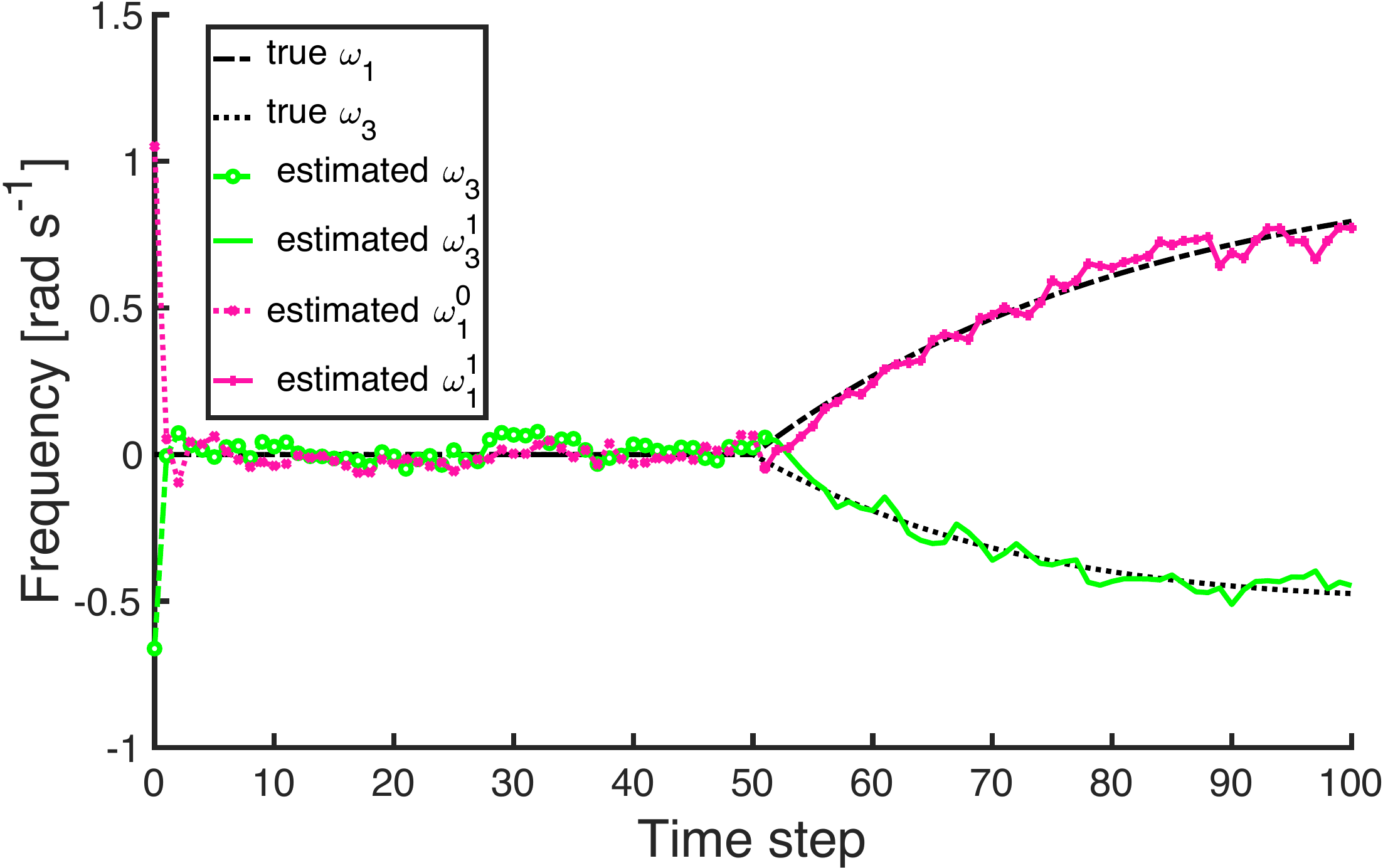}
		\caption{Estimated vs \textit{true} trajectory of frequencies $\omega_1$ and $\omega_3$.}
		\label{fig:omega}
	\end{center}
\end{figure} 
%\vspace{.5cm}
\begin{figure}[tb]
	\begin{center}
		\includegraphics[width=.8\columnwidth]{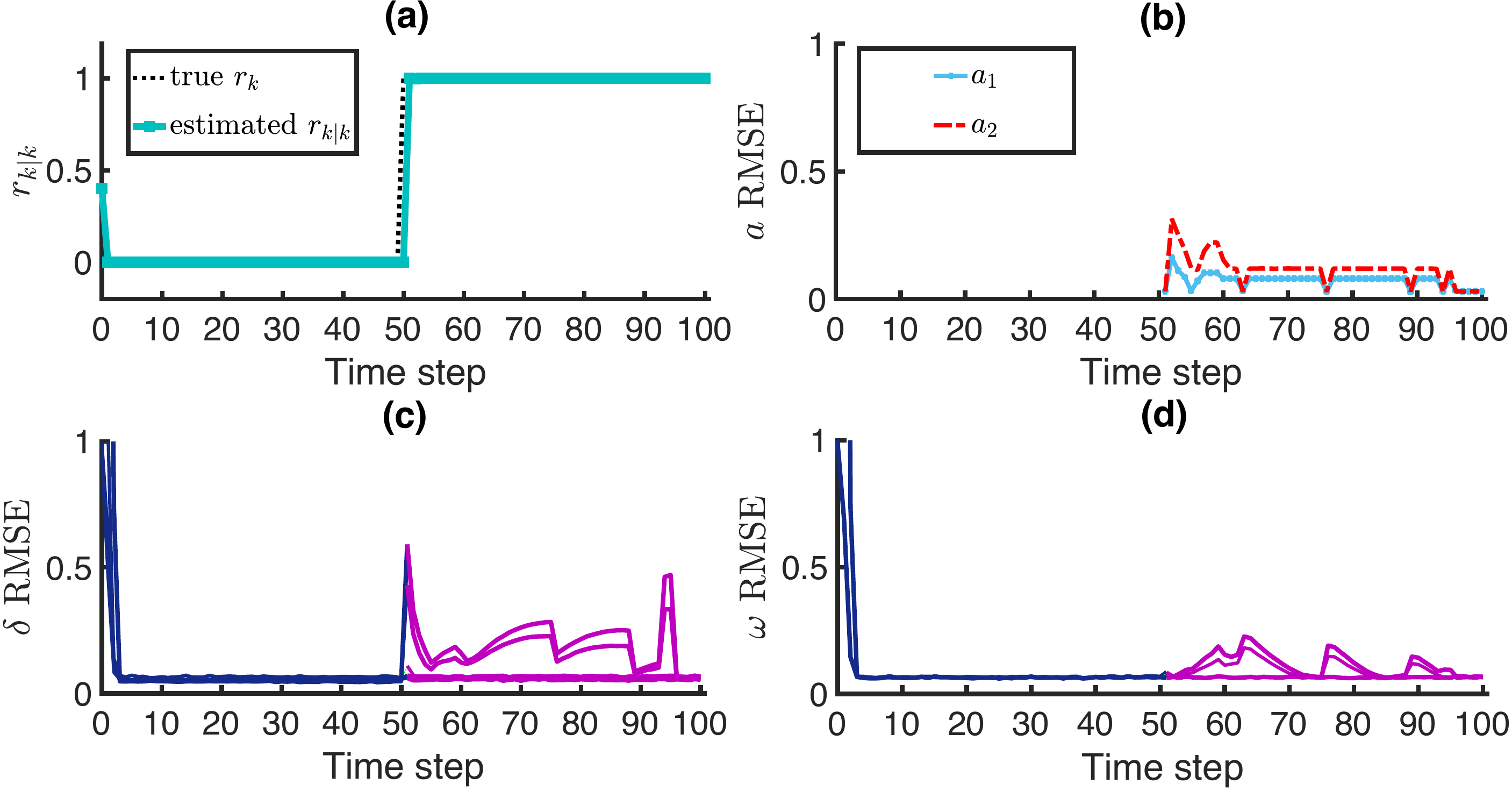}
		\caption{Performance of the GM-HBF under packet substitution attack ($p_f=0.3$) in terms of 
			(a) attack detection, (b) attack reconstruction, and (c)-(d) state estimation.}
		\label{fig:pf1}
	\end{center}
\end{figure} 
\begin{figure}[tb]
	\begin{center}
		\includegraphics[width=.6\columnwidth]{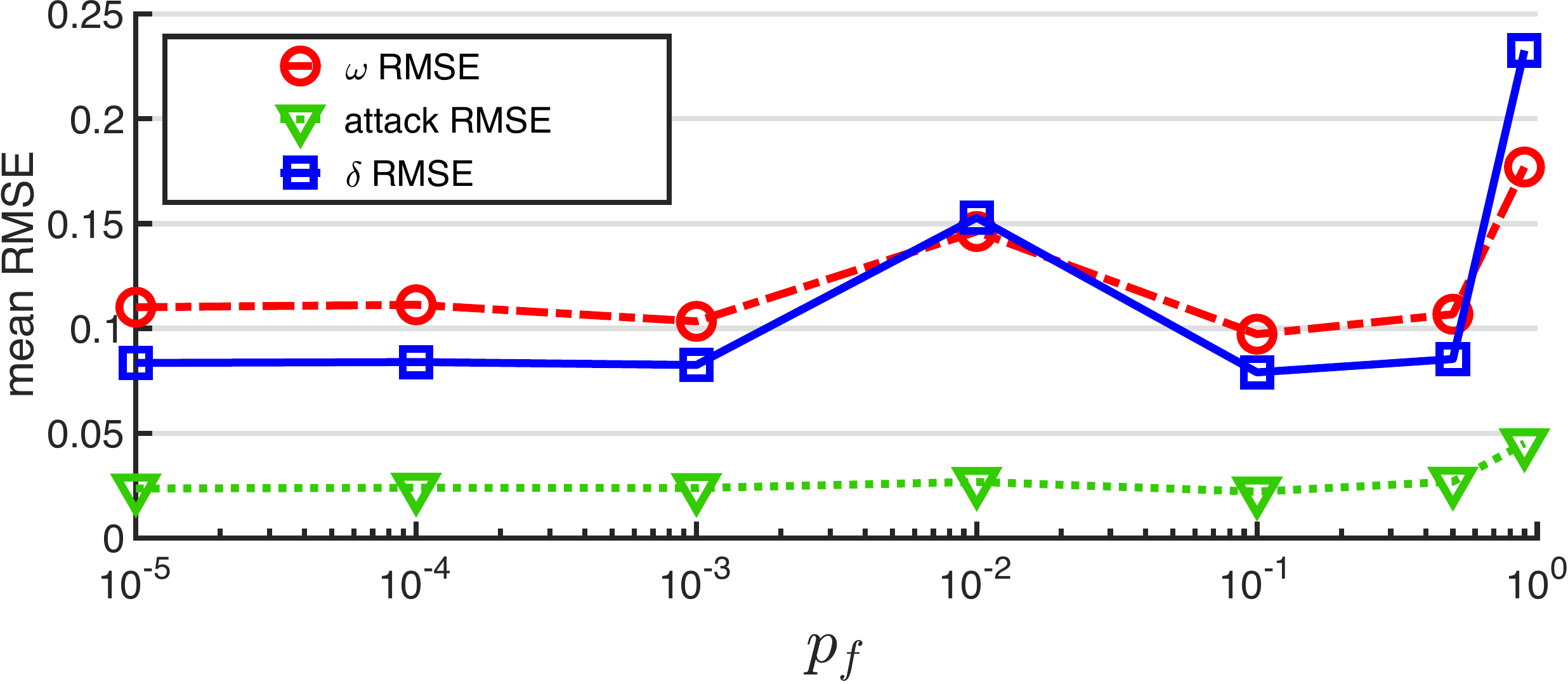}
		\caption{Mean RMSE on state (generators' rotor angles and frequencies) and attack estimation under packet substitution attack as a function of filter's parameter $p_f$. Simulated packet substitutions occur with probability $\bar{p}_f=0.1$. 
			The choice of $p_f$ can improve estimation performance (the best results are obtained when $p_f=\bar{p}_f$) which, however, turns out to be comparable for most parameter's values in the set $\{ 10^{-5}, 10^{-4}, 10^{-3}, 10^{-2}, 0.1, 0.5, 0.9 \}$.
		}
		\label{fig:pf2}
	\end{center}
\end{figure} 

Finally, Fig.~\ref{fig:omega} shows the performance of the GM-HBF in estimating the generator frequencies $\omega_1$ and $\omega_3$, before and after the appearance of the signal attack on the victim loads.
The performance of the proposed GM-HBF under packet substitution attack, i.e. the filter adopting the correction step described in part 1) of Section~\ref{sec3.1}, is shown in Fig.~\ref{fig:pf1} for $p_f=0.3$ and $p_d=1$. 
It is worth noting that the probability of packet substitution $p_f$ can be seen as a design parameter which can be suitably tuned so as to enhance estimation performance. 
This is illustrated in Fig.~\ref{fig:pf2} where the mean (over time, components and Monte Carlo runs) RMSE on state/attack estimation is shown as a function of parameter $p_f$. By contrast, simulation results indicated that the choice on $p_f$ does not significantly affect the overall attack detection performance.

\section{Conclusions}\label{sec7}
This paper proposed a general framework to solve 
resilient state estimation for (linear/nonlinear) cyber-physical systems considering switching signal attacks, fake measurement injection and packet substitution.
Random finite sets have been exploited in order to model the switching nature of the signal attack as well as the possible presence of fake measurements, and a Bayesian random set estimation problem has been formulated for jointly detecting a signal attack and estimating the system state.
In this way, a hybrid Bernoulli filter for the Bayes-optimal solution of the posed problem has been derived and implemented as a Gaussian-sum filter. 
%In the present work, the simultaneous attack detection and secure state estimation problem is addressed from the estimator's perspective, 
%i.e. without modeling any strategic interaction between estimator and attacker which, by contrast, characterizes game-theoretic (worst-case) approaches.
Numerical examples concerning both a benchmark system with direct feedthrough and a realistic energy delivery system have been presented so as to demonstrate the potentials 
and the real-world applicability
of the proposed approach.
Future work will concern worst-case performance degradation analysis for the developed filter and its application to resilient state estimation in distributed settings with non-secure communication links. 

\vspace{.5cm}

%The use of the \LaTeX\ cross-reference system
%for figures, tables, equations, etc., is encouraged
%(using \verb"\ref{<name>}" and \verb"\label{<name>}").

%\subsection{Acknowledgements} An Acknowledgements section is started with \verb"\ack" or
%\verb"\acks" for \textit{Acknowledgement} or
%\textit{Acknowledgements}, respectively. It must be placed just
%before the References.

%\subsection{Bibliography}

\bibliographystyle{IEEEtr}
\bibliography{bib_special_issue}
 
%\bibliographystyle{NJDnatbib}

%\clearpage

\end{document}